\documentclass[dvips,envcountsect,envcountsame]{llncs}
\usepackage{macros}
\usepackage{psfrag}
\usepackage{times}
\usepackage{tikz}
\usepackage{mathtools}
\usepackage{graphicx}
\usepackage{ifthen}
\newcommand{\techRep}{true}
\newcommand{\iftechrep}{\ifthenelse{\equal{\techRep}{true}}}

\allowdisplaybreaks[4]

\title{Space-efficient scheduling of \\ stochastically generated tasks}

\begin{document}

\author{Tom\'{a}\v{s} Br\'{a}zdil\inst{1}\thanks{Supported by Czech Science Foundation, grant No.{} P202/10/1469.}
 \and Javier Esparza\inst{2} \and Stefan Kiefer\inst{3}\thanks{Supported by the EPSRC project \emph{Automated Verification of Probabilistic Programs}.}
 \and Michael Luttenberger\inst{2}}

\institute{Faculty of Informatics, Masaryk University, Brno, Czech Republic
 \and Institut f{\"u}r Informatik, Technische Universit{\"a}t M{\"u}nchen, Germany
 \and Oxford University Computing Laboratory, UK
}



\sloppy

\maketitle

\begin{abstract}
We study the problem of scheduling tasks for execution by a processor
when the tasks can stochastically generate new tasks. Tasks can be of
different types, and each type has a fixed, known probability of generating
other tasks.
We present results on
the random variable $S^\sigma$ modeling the maximal space needed by the
processor to store the currently active tasks when acting under the
scheduler~$\sigma$. We obtain tail bounds for the distribution of $S^\sigma$ for
both offline and online schedulers, and investigate
the expected value $\Ex{S^\sigma}$.
\end{abstract}

\section{Introduction}

We study the problem of scheduling tasks that can stochastically generate
new tasks. We assume that the execution of a task $\tau$ can generate a set of
subtasks. Tasks can be of different types,
and each type has a fixed, known probability of generating new subtasks.
\iftechrep{\par}{}
Systems of tasks can be described using a notation
similar to that of stochastic grammars. For instance
\iftechrep{}{\vspace{-2mm}}
\[
 \begin{array}{cc}
   X  \btran{0.2} \langle X, X \rangle  \qquad X \btran{0.3} \langle X, Y \rangle \qquad  X \btran{0.5} \emptyset \qquad 
   Y  \btran{0.7} \langle X \rangle \qquad Y \btran{0.3} \langle Y \rangle
 \end{array}
\iftechrep{}{\vspace{-1mm}}
\]
\noindent describes a system with two types of tasks.
Tasks of type~$X$ can generate $2$ tasks of type~$X$, one task of each type,
or zero tasks with probabilities
$0.2$, $0.3$, and  $0.5$, respectively (angular brackets denote multisets).
Tasks of type $Y$ can generate
one task, of type $X$ or $Y$, with probability $0.7$ and $0.3$.
Tasks are executed by one processor. The processor repeatedly
selects a task from a pool of unprocessed tasks, processes it, and puts
the generated subtasks (if any) back into the pool. The pool initially
contains one task of type $X_0$, and the next task to be processed is
selected by a {\em scheduler}.

We study random variables modeling the time and space
needed to {\em completely} execute a task $\tau$, i.e., to empty the pool of unprocessed tasks
assuming that initially the pool only contains task $\tau$.
We assume that processing a task takes one time unit, and
storing it in the pool takes a unit of memory. So the {\em completion time}
is given by the total number of tasks processed, and the {\em completion space} by
the maximum size reached by the pool during the computation.
The completion time has been studied in \cite{EKM:prob-PDA-expectations},
and so the bulk of the paper is devoted to studying the
distribution of the completion space for different classes of schedulers.

Our computational model is abstract, but relevant for different scenarios.
In the context of search problems, a task is a
problem instance, and the 
scheduler is part of a branch-and-bound algorithm
(see e.g.~\cite{KarpZhang93}). In the more general
context of multithreaded computations, a task models a thread, which
may generate new threads.
The problem of scheduling multithreaded computations space-efficiently
on {\em multi}processor machines has been extensively studied
(see e.g. \cite{NarBel99,BluLei99,AroBluPla02,ALHH08}). These papers
assume that schedulers know nothing about the program, while we consider
the case in which stochastic information on the program behaviour is available
(obtained from sampling).

We study the performance of {\em online} schedulers that know only the past of
the computation, and compare them with
the {\em optimal offline} scheduler, which has complete information about
the future. Intuitively, this scheduler has access to an oracle that knows
how the stochastic choices will be resolved. The oracle can be replaced by a machine
that inspects the code of a task and determines which subtasks it will generate (if any).

We consider task systems with completion probability 1, which
can be further divided into those with finite and infinite expected
completion time, often called {\em subcritical} and {\em critical}.
Many of our results are related to the probability generating functions (pgfs)
associated to a task system. The functions for the example above are
$f_X(x,y)  =  0.2 x^2 + 0.3 xy + 0.5$ and $f_Y(x,y)  =  0.7 x + 0.3 y$,
and the reader can easily guess the formal definition. The completion
probability is the least
fixed point of the system of pgfs~\cite{Harris63}.

Our first results (Section \ref{sec:optimal}) concern the distribution of the completion space $\xo$ of the optimal
offline scheduler ${\it op}$\, on a fixed but arbitrary task system with $\vf(\vx)$ as pgfs (in vector form).
We exhibit a very surprising connection between the probabilities
$\pr{\xo = k}$ and the {\em Newton approximants} to the least fixed point of $\vf(\vx)$
(the approximations to the least fixed point obtained by applying Newton's method for
approximating a zero of a differentiable function to $\vf(\vx)-\vx = \vzero$ with seed $\vzero$).
This connection allows us to apply recent results on the
convergence speed of Newton's method \cite{KLE07:stoc,EKL08:stacs}, leading to
tail bounds of~$\xo$, i.e., bounds on $\pr{\xo \geq k}$.
We then study (Section \ref{sec:online}) the
distribution of $S^\sigma$ for an online scheduler $\sigma$, and obtain upper and lower
bounds for the
performance of {\em any} online scheduler in subcritical systems. These bounds
suggest a way of assigning weights to task types reflecting how likely they are
to require large space. We study {\em light-first} schedulers, in which ``light''
tasks are chosen before
``heavy'' tasks with larger components, and obtain an improved tail bound.

So far we have assumed that there are no dependencies between tasks, requiring a task to be executed
before another. We study in Section \ref{sub:depth-first} the case in which a task can only terminate
after all the tasks it has (recursively) spawned have terminated. These are the {\em strict}
computations studied in \cite{BluLei99}. The optimal scheduler in this case is the {\em depth-first} scheduler,
 i.e., the one that
completely executes the child task before its parent,
resulting in the familiar stack-based execution. Under this scheduler our tasks are
equivalent to special classes of recursive state machines \cite{EYstacs05Extended} and
probabilistic pushdown automata \cite{EKM04}. We determine the exact asymptotic performance
of depth-first schedulers, hereby making use of recent results~\cite{BEK09:fsttcs}.

We restrict
ourselves to the case in which a task has at most two children,
i.e., all rules  $X  \btran{p} \langle X_1, \ldots, X_n \rangle$
satisfy $n \leq 2$. This case already allows to model
the forking-mechanism underlying many multithreaded operating systems,
e.g. Unix-like systems.

\emph{Related work.}
Space-efficient scheduling for search problems
or multithreaded computations has been studied in
\cite{KarpZhang93,NarBel99,BluLei99,AroBluPla02,ALHH08}. These papers assume that
nothing is known about the program generating
the computations. We study the case in which statistical information is available
on the probability that computations split or die.
\iftechrep{\par}{}
The theory of {\em branching processes} studies stochastic processes modeling
populations whose members can reproduce or die \cite{Harris63,AthreyaNey:book}.
In computer science terminology, all existing work on branching processes
assumes that the number of processors is {\em unbounded}
\cite{Athreya88,BorVat96,Lind76,Nerm77,Pakes98,Spataru91}.
\iftechrep{We study the 1-processor case, and to our knowledge we are the first to do so.}{To our knowledge, we are the first to study the 1-processor case.}

\emph{Structure of the paper.}
The rest of the paper is structured as follows.
The preliminaries in Section~\ref{sec:prelim} formalize the notions from the introduction
 and summarize known results on which we build.
In Section~\ref{sec:optimal} we study \iftechrep{the performance of} optimal offline schedulers.
Section~\ref{sec:online} is dedicated to online schedulers.
First we prove performance bounds that hold uniformly for all online schedulers,
 then we prove improved bounds for light-first schedulers,
 and finally we determine the exact asymptotic behaviour of depth-first schedulers.
In Section~\ref{sec:expectations} we obtain several results on the expected space consumption
 under different schedulers.
Section~\ref{sec:conclusions} contains some conclusions.
Full proofs can be found in \iftechrep{the appendix.}{\cite{BEKL10:IcalpTechRep}}.

\section{Preliminaries} \label{sec:prelim}
Let $A$ be a finite set.
We regard elements of $\Nat^A$ and $\R^A$ as {\em vectors} and use boldface
(like $\vu, \vv$) to denote vectors.
The vector whose components are all~$0$ (resp.~$1$) is denoted by~$\vzero$ (resp.~$\vone$).
We use angular brackets to denote multisets and often identify multisets over $A$ and
vectors indexed by~$A$. For instance, if $A = \{X,Y\}$ and
$\vv \in \Nat^A$ with $\vv_X = 1$ and $\vv_Y = 2$,
then $\vv = \langle X,Y,Y \rangle$. We often shorten $\langle a \rangle$ to~$a$.
$M_A^{\leq 2}$ denotes the multisets over $A$ containing at most $2$ elements.

\begin{definition}
A {\em task system} is a tuple $\Delta=(\Gamma,\btran{},\Prob, \Init)$ where
$\Gamma$ is a finite set of \emph{task types}, ${\btran{}}\subseteq \Gamma\times M_\Gamma^{\leq 2}$ is a set of \emph{transition rules},
 $\Prob$ is a function assigning positive probabilities to transition rules so that for every $X\in \Gamma$ we have $\sum_{X\tran{} \alpha} \Prob((X,\alpha))=1$,
 and $\Init \in \Gamma$ is the {\em initial type}.
\end{definition}
We write $X\btran{p} \alpha$ whenever $X\btran{}\alpha$ and $\Prob((X,\alpha))=p$.
Executions of a task system are modeled as family trees, defined as follows.
Fix an arbitrary total order $\preceq$ on $\Gamma$. A
{\em family tree} $t$ is a pair $(N,L)$ where $N\subseteq \{0,1\}^*$ is a
finite binary tree (i.e.\ a prefix-closed finite set of words over $\{0,1\}$)
and $L:N\tran{} \Gamma$ is a labelling such that every node $w\in N$ satisfies
one of the following conditions: $w$ is a leaf and $L(w)\btran{} \varepsilon$, or
$w$ has a unique child $w0$, and $L(w)$ satisfies $L(w)\btran{} L(w0)$, or
$w$ has two children $w0$ and $w1$, and $L(w0)$, $L(w1)$ satisfy
$L(w)\btran{} \langle L(w0), L(w1) \rangle$ and $L(w0)\preceq L(w1)$.
Given a node $w\in N$, the subtree of $t$ rooted at $w$, denoted by ${t}_{w}$,
is the family tree $(N',L')$ such that $w'\in N'$ iff $ww'\in N$ and
$L'(w')=L(ww')$ for every $w'\in N'$.
If a tree~$t$ has a subtree $t_0$ or $t_1$, we call this subtree a {\em child} of~$t$.
(So, the term {\em child} can refer to a node or a tree, but there will be no confusion.)

We define a function $\Pr$ which, loosely speaking, assigns to a family tree $t=(N,L)$
its probability (see the assumption below).
Assume that the root of $t$ is labeled by $X$.
If $t$ consists only of the root, and
$X \btran{p} \varepsilon$, then $\pr{t} = p$; if the root has only one child (the node~$0$)
labeled by $Y$, and $X \btran{p} Y$, then $\pr{t}=p \cdot \pr{t_0}$; if the root has two children
(the nodes $0$ and~$1$) labeled by $Y$ and $Z$, and $X \btran{p} \langle Y, Z \rangle$,
then $\pr{t}=p \cdot \pr{{t}_{0}} \cdot \pr{{t}_{1}}$.
We denote by ${\mathcal T}_X$ the set of all family trees whose root is labeled by $X$, and 
by ${\Pr}_X$ the restriction of $\Pr$ to ${\mathcal T}_X$.
We drop the subscript of~${\Pr}_X$ if $X$ is understood.

\vspace{-2mm}
\begin{example}
Figure \ref{fig:ex-1} shows (a) a task system
with $\Gamma=\{X,Y,Z\}$; 
and (b) a family tree $t$ of the system with probability $\pr{t} = 0.25 \cdot 0.1 \cdot 0.75 \cdot 0.6 \cdot 0.4 \cdot 0.9$.
The name and label of a node are written close to it.

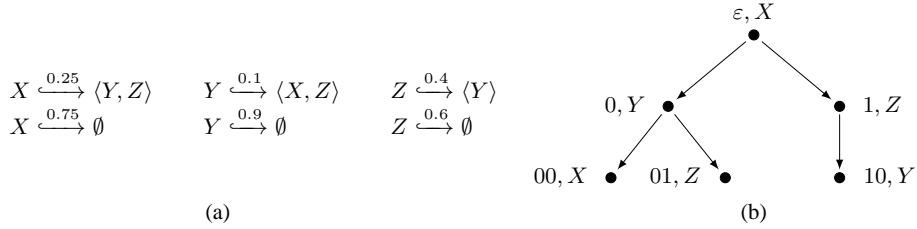
\begin{figure}
\iftechrep{}{\vspace{-3mm}}
\scalebox{0.95}{
\begin{tikzpicture}

\begin{scope}
[shift={(1,0)}]
\node (a) at (3.0,-2.5) {(a)};

\node (text) at (3.5,-1.0) {$\begin{array}{lcl@{\hspace{7mm}}lcl@{\hspace{7mm}}lcl}
X & \xhookrightarrow{0.25} & \langle Y,Z \rangle & Y & \xhookrightarrow{0.1} & \langle X,Z \rangle & Z & \xhookrightarrow{0.4} & \langle Y \rangle \\
X & \xhookrightarrow{0.75} & \emptyset           & Y & \xhookrightarrow{0.9} & \emptyset           & Z & \xhookrightarrow{0.6} & \emptyset
\end{array}$};
\end{scope}

\begin{scope}[-latex,node distance=3mm,shift={(11.5,0)}]
  \node (b) at (0,-2.5) {(b)};

  \node (0) at (0,0) {};
  \node (1) at (-1.2,-1) {};
  \node (2) at (1.2,-1) {};
  \node (11) at (-2,-2) {};
  \node (12) at (-0.4,-2) {};
  \node (21) at (1.2,-2) {};

  \foreach \n in {0,1,2,11,12,21}
  {
    \draw[fill = black] (\n) circle (2pt);
  }

  \draw (0) -- (1);
  \draw (0) -- (2);
  \draw (1) -- (11);
  \draw (1) -- (12);
  \draw (2) -- (21);


  \node[above of=0]{$\ew,X$};
  \node[left of=1,  node distance=6mm]{$0,Y$};
  \node[right of=2, node distance=6mm]{$1,Z$};
  \node[left of=11, node distance=7mm]{$00,X$};
  \node[left of=12, node distance=7mm]{$01,Z$};
  \node[right of=21, node distance=7mm]{$10,Y$};
\end{scope}

\end{tikzpicture}
}
\caption{(a) A task system. (b) A family tree.}
\label{fig:ex-1}
\iftechrep{}{\vspace{-3mm}}
\end{figure}
\end{example}
\emph{Assumptions.} Throughout the paper we assume that a task system
\mbox{$\Delta=(\Gamma,\btran{},\Prob, \Init)$}
satisfies the following two conditions for every type $X \in \Gamma$: (1) $X$ is
{\it reachable} from $\Init$, meaning that some tree in ${\mathcal T}_\Init$ contains a node
labeled by $X$, and (2) $\pr{{\mathcal T}_X} = \sum_{t\in {\mathcal T}_X} \pr{t} =1$. So
we assume that $({\mathcal T}_X,{\Pr}_X)$ is
a discrete probability space with ${\mathcal T}_X$ as set of elementary events and
${\Pr}_X$ as probability function. This is the formal counterpart to
assuming that every task is completed with probability~1.

\begin{proposition} \label{prop:prelim-assumptions}
 It can be decided in polynomial time whether assumptions (1) and~(2) are satisfied.
\end{proposition}
\begin{proof}
(1) is trivial.
For (2) let the {\em probability generating function} (pgf) of the task system be defined
 as the function $\vf : \R^\Gamma \to \R^\Gamma$ of $\Delta$ where for every $X \in \Gamma$
\iftechrep{}{\vspace{-1mm}}
 \[ \vf_X(\vv) = \sum_{X \btran{p} \langle Y,Z\rangle}
 p \cdot \vv_Y \cdot \vv_Z + \sum_{X \btran{p} \langle Y\rangle} p \cdot
 \vv_Y + \sum_{X \btran{p} \emptyset} p \,.
\iftechrep{}{\vspace{-2mm}}
 \]
It is well known (see e.g.~\cite{Harris63}) that (2) holds iff the least
nonnegative fixed point of~$\vf$ equals~$\vone$, which is decidable in polynomial time~\cite{EYstacs05Extended}.
\qed
\end{proof}

\emph{Derivations and schedulers.} Let $t=(N,L)$ be a family tree. A {\em state} of $t$ is a
maximal subset of $N$ in which no node is a proper prefix of another node (graphically, no node
is a proper descendant of another node). The elements of a state
are called {\em tasks}. If $s$ is a state and $w \in s$, then the {\em $w$-successor of $s$}
is the uniquely determined state $s'$ defined as follows:
if $w$ is a leaf of $N$, then $s' = s \setminus \{ w\}$; if $w$ has
one child $w0$, then $s' = (s \setminus \{ w\})\cup \{w0\}$; if
$w$ has two children $w0$ and $w1$, then $s' = (s \setminus \{ w\}) \cup \{w0,w1\}$.
We write $s \Rightarrow s'$ if $s'$ is the $w$-successor of $s$ for some $w$.
A \emph{derivation of~$t$} is a sequence
$s_1 \Rightarrow \ldots \Rightarrow s_k$ of states such that $s_1 = \{\epsilon\}$ and $s_k = \emptyset$.
A {\em scheduler} is a mapping $\sigma$ that assigns to
a family tree $t$ a derivation $\sigma(t)$ of $t$.
If $\sigma(t) = (s_1 \Rightarrow  \ldots \Rightarrow s_k)$,
then for every $1\leq i<k$ we denote by $\sigma(t)[i]$ a task of $s_i$
such that $s_{i+1}$ is the $\sigma(t)[i]$-successor of~$s_i$. Intuitively,
$\sigma(t)[i]$ is the task of $s_i$ scheduled by $\sigma$. This definition allows for
schedulers that know the tree, and so how future
tasks will behave. In Section~\ref{sec:online} we define and study online
schedulers which only know the past of the computation.
Notice that schedulers are deterministic (non-randomized).

\vspace{-2mm}
\begin{example} \label{ex:derivations}
A scheduler~$\sigma_1$ may schedule the tree $t$ in Figure~\ref{fig:ex-1} as follows:
 $\{\ew\}\Rightarrow \{0,1\} \Rightarrow \{0,10\} \Rightarrow \{0 \} \Rightarrow \{00,01\} \Rightarrow \{01\} \Rightarrow \{\}$.
Let $\sigma_2$ be the scheduler which always picks the least unprocessed task  w.r.t.\ the lexicographical order on $\{0,1\}^\ast$.
(This is an example of an online scheduler.)
It schedules~$t$ as follows:
 \mbox{$\{\ew\} \Rightarrow \{0,1\} \Rightarrow \{00,01,1\}\Rightarrow \{01,1\}\Rightarrow \{1\}\Rightarrow \{10\} \Rightarrow \{\}$.}
\end{example}

\emph{Time and space.} Given $X\in \Gamma$, we define a random variable
$T_X$, the {\em completion time of $X$}, that assigns to a tree $t \in {\mathcal T}_X$ its number of nodes.
Assuming that tasks are executed for one time unit before its generated subtasks
are returned to the pool, $T_X$ corresponds to the time required to completely execute $X$.
Our assumption~(2) guarantees that $T_X$ is finite with probability~$1$, but its expectation
$\Ex{T_X}$ may or may not be finite.
A task system $\Delta$ is called {\em subcritical} if $\Ex{T_X}$ is finite
for every $X\in \Gamma$. Otherwise it is called {\em critical}. If $\Delta$ is subcritical,
then $\Ex{T_X}$ can be easily computed by solving a system of linear equations
\cite{EKM:prob-PDA-expectations}. The notion of criticality comes from the theory of branching processes,
see e.g.~\cite{Harris63,AthreyaNey:book}. Here we only recall the following results:

\begin{proposition}[\cite{Harris63,EYstacs05Extended}] \label{prop:critical-spectral}
Let $\Delta$ be a task system with pgf~$\vf$.
Denote by $\vf'(\vone)$ the Jacobian matrix of partial derivatives of~$\vf$ evaluated at~$\vone$.
If $\Delta$ is critical, then the spectral radius of $\vf'(\vone)$ is equal to~$1$;
otherwise it is strictly less than~$1$.
It can be decided in polynomial time whether $\Delta$ is critical.
\end{proposition}
A state models a pool of tasks awaiting to be scheduled.
We are interested in the maximal size of the pool during the execution of a derivation.
So we define the random {\em completion space} $S^\sigma_X$ as follows.
If \mbox{$\sigma(t) = (s_1 \Rightarrow \ldots \Rightarrow s_k)$}, then
 $S^\sigma_X(t) := \max\{|s_1|, \ldots, |s_k|\}$, where $|s_i|$ is the cardinality of $s_i$.
Sometimes we write $S^{\sigma}(t)$, meaning
$S^{\sigma}_X(t)$ for the type $X$ labelling the root of $t$.
If we write $S^{\sigma}$ without specifying the application to any tree,
then we mean $S^{\sigma}_{\Init}$.

\vspace{-2mm}
\begin{example}
 For the schedulers of Example~\ref{ex:derivations} we have $S^{\sigma_1}(t) = 2$ and $S^{\sigma_2}(t) = 3$.
\end{example}

\section{Optimal (Off\/line) Schedulers}
\label{sec:optimal}

Let $\xo$ be the random variable that assigns to a family tree the minimal completion space
of its derivations. We call $\xo(t)$ the {\em optimal completion space} of~$t$. The
optimal scheduler assigns to each tree a derivation with optimal completion space.
In the multithreading scenario, it corresponds to a scheduler that
can inspect the code of a thread and decide whether it will
spawn a new thread or not.
%
Note that, although the optimal scheduler ``knows'' how the stochastic choices are resolved,
 the optimal completion space $\xo(t)$ is still a random variable, because it depends
 on a random tree.
The following proposition characterizes the optimal completion space of a tree in
 terms of the optimal completion space of its children.
\begin{proposition} \label{prop:char-optimal}
 Let $t$ be a family tree.
 Then
 \[
  \xo(t) =
   \begin{cases}
    \min\left\{ 
                \begin{array}{l}
                  \max\{\xo(t_0)+1, \xo(t_1)\}, \\
                  \max\{\xo(t_0), \xo(t_1)+1\}
                \end{array}
        \right\}
             & \text{if $t$ has two children $t_0$, $t_1$} \\
    \xo(t_0) & \text{if $t$ has exactly one child $t_0$} \\
    1        & \text{if $t$ has no children.}
   \end{cases}
 \]
\end{proposition}
{\em Proof sketch.}
 The only nontrivial case is when $t$ has two children $t_0$ and~$t_1$.
 Consider the following schedulings for $t$, where $i \in \{0,1\}$: Execute first all
 tasks of $t_i$ and then all tasks of $t_{1-i}$; within both $t_i$ and $t_{1-i}$, execute
 tasks in optimal order. While executing $t_i$, the root task
 of~$t_{1-i}$ remains in the pool, and so the completion space is
 $s(i) = \max\{\xo(t_i)+1, \xo(t_{1-i})\}$. The optimal scheduler chooses the value of $i$
 that minimizes $s(i)$.
\qed

\vspace{2mm}

Given a type~$X$, we are interested in the probabilities $\pr{\xo_X \leq k}$ for $k \geq 1$.
Proposition~\ref{prop:char-optimal} yields a recurrence
relation which at first sight seems difficult to handle. However, using
results of ~\cite{EKL07:stacs,EKL07:dlt} we can exhibit a surprising
connection between these probabilities and the pgf $\vf$.

Let $\vmu$ denote the least fixed point of~$\vf$ and recall from the proof of Proposition~\ref{prop:prelim-assumptions} that $\vmu = \vone$.
Clearly, $\vone$ is a zero of $\vf(\vx) - \vx$. It has recently been shown
that $\vmu$ can be computed by applying to $\vf(\vx) - \vx$ Newton's method for
approximating a zero of a differentiable function \cite{EYstacs05Extended,KLE07:stoc}.
More precisely, $\vmu = \lim_{k\rightarrow \infty} \ns{k}$ where
\begin{equation*}
\ns{0} = \vzero \quad \text{ and } \quad \ns{k+1} = \ns{k} +
(\Id - \vf'(\ns{k}))^{-1} \left( \vf(\ns{k}) - \ns{k} \right)\;
\end{equation*}
\noindent and $\vf'(\ns{k})$ denotes the Jacobian matrix of partial derivatives of~$\vf$ evaluated at~$\ns{k}$
 and $\Id$ the identity matrix.
Computing $\vmu$, however, is in our case uninteresting, because we already know that $\vmu = \vone$.
So, why do we need Newton's method? Because
the sequence of Newton approximants provides exactly the information we are
looking for:

\begin{theorem} \label{thm:whitebox}
 $\pr{\xo_X \leq k} = \ns{k}_X$ for every type $X$ and every $k \geq 0$.
\end{theorem}
\newcommand{\mathcalT}[1]{{\mathcal T}_{#1}}
\newcommand{\mathcalB}[2]{{\mathcal B}_{#1}^{(#2)}}
{\em Proof sketch.}
 We illustrate the proof idea on the one-type task system with pgf
 $f(x) = px^2 + q$, where $q=1-p$.
 Let $\mathcalT{\leq k}$ and $\mathcalT{=k}$ denote the sets of trees~$t$ with $\xo(t) \le k$
 and $\xo(t) = k$, respectively.
 We show $\pr{\mathcalT{\leq k}} = \nsc{k}$ for all~$k$ by induction on $k$. The case $k=0$ is trivial.
 Assume that $\nsc{k} = \pr{\mathcalT{\leq k}}$ holds for some~$k \geq 0$. We prove
 $\pr{\mathcalT{\leq k+1}} = \nsc{k+1}$. Notice that

 $\nsc{k+1} :=  \nsc{k}+\frac{f(\nsc{k}) - \nsc{k}}{1- f'(\nsc{k})} =
 \nsc{k} + (f(\nsc{k}) - \nsc{k}) \cdot \sum_{i=0}^\infty f'(\nsc{k})^i.$

 \noindent Let $\mathcalB{k+1}{0}$ be the set of trees that have two children
 both of which belong to $\mathcalT{=k}$, and, for every $i \ge 0$, let $\mathcalB{k+1}{i+1}$ be the set of
 trees with two children, one belonging to~$\mathcalT{\leq k}$, the other one
 to~$\mathcalB{k+1}{i}$.
 By Proposition~\ref{prop:char-optimal} we have $\mathcalT{\leq k+1} = \bigcup_{i\geq 0}\mathcalB{k+1}{i}$.
 We prove $\pr{\mathcalB{k+1}{i}}=f'(\nsc{k})^i~(f(\nsc{k}-\nsc{k})$ by an (inner) induction on~$i$,
 which completes the proof. For the base $i=0$, let ${\mathcal A}_{\leq k}$ be the set
 of trees with two children in $\mathcalT{\leq k}$; by induction hypothesis we have
 $\pr{{\mathcal A}_{\leq k}} = p \nsc{k} \nsc{k}$. In a tree of ${\mathcal A}_{\leq k}$ either
 (a) both children belong to $\mathcalT{=k}$, and so $t \in \mathcalB{k+1}{0}$,
 or (b) at most one child belongs to $\mathcalT{=k}$. By Proposition~\ref{prop:char-optimal},
 the trees satisfying~(b) belong to~$\mathcalT{\leq k}$. In fact, a stronger property holds:
 a tree of~$\mathcalT{\leq k}$ either satisfies~(b) or it has one single node.
 Since the probability of the tree with one node is~$q$, we get
 $\pr{{\mathcal A}_{\leq k}} = \pr{\mathcalB{k+1}{0}} + \pr{\mathcalT{\leq k}} - q$.
 Applying the induction hypothesis again we obtain
 $\pr{\mathcalB{k+1}0} = p \nsc{k} \nsc{k} + q - \nsc{k} = f(\nsc{k}) - \nsc{k}$.
 For the induction step, let $i > 0$. Divide $\mathcalB{k+1}{i}$ into two sets, one
 containing the trees whose
 left (right) child belongs to~$\mathcalB{k+1}{i}$ (to $\mathcalT{\leq k}$), and the other
 the trees whose left (right) child belongs to~$\mathcalT{\leq k}$ (to $\mathcalB{k+1}{i}$).
 Using both induction hypotheses, we get that the probability of each set is
 $p \nsc{k} f'(\nsc{k})^i
 (f(\nsc{k}) - \nsc{k})$. So
 $\pr{\mathcalB{k+1}{i+1}} = (2p \nsc{k}) \cdot f'(\nsc{k})^i  (f(\nsc{k}) - \nsc{k})$.
 Since $f(x) = px^2 + q$ we have $f'(\nsc{k}) = 2p \nsc{k}$, and so $\pr{\mathcalB{k+1}{i+1}} =
 f'(\nsc{k})^{i+1} (f(\nsc{k} - \nsc{k})$ as desired.
\qed

\vspace{-2mm}
\begin{example}
Consider the task system $X \btran{p} \langle X, X \rangle, \ X \btran{q} \emptyset$
with pgf $f(x) =  px^2 + q$, where $p$ is a parameter and $q = 1-p$. The least
fixed point of $f$ is $1$ if $p \leq 1/2$ and $q/p$ otherwise. So we consider only the case
$p \leq 1/2$. The system is critical for $p = 1/2$ and subcritical
for $p < 1/2$. Using Newton approximants we obtain the following recurrence relation for the
distribution of the optimal scheduler, where $p_k := \pr{\xo \geq k} = 1 - \nsc{k-1}$:
$p_{k+1} = (p p_k^2)/(1-2p + 2p p_k)$. In particular, for the critical value $p=1/2$ we get $p_k = 2^{1-k}$ and
$\Ex{\xo} =  \sum_{k \ge 1} \pr{\xo \geq k} = 2$.
\end{example}

Theorem \ref{thm:whitebox} allows to compute the probability mass function of $\xo$.
As a Newton iteration requires $\bigo(|\Gamma|^3)$ arithmetical operations,
 we obtain the following corollary, where by the unit
 cost model we refer to the cost in the Blum-Shub-Smale model, in which
 arithmetic operations have cost 1 independently of the size of the operands \cite{BlumShubSmale:book}.
\begin{corollary} \label{cor:optimal-distribution-cost}
$\pr{\xo_X = k}$ can be computed in time $\bigo(k \cdot |\Gamma|^3)$ in the unit cost model.
\end{corollary}

\noindent It is easy to see that Newton's method converges quadratically for
subcritical systems (see e.g. \cite{OrtegaRheinboldt:book}).
For critical systems, it has recently been proved that Newton's method still converges linearly
\cite{KLE07:stoc,EKL08:stacs}. These results lead to tail bounds for~$\xo_X$:

\begin{corollary} \label{cor:optimal-expectation}
 For any task system~$\Delta$ there are real numbers $c > 0$ and $0 < d < 1$ such that $\pr{\xo_X \ge k} \le c \cdot d^k$ for all $k \in \Nat$.
 If $\Delta$ is subcritical, then there are real numbers $c > 0$ and $0 < d < 1$ such that $\pr{\xo_X \ge k} \le c \cdot d^{2^k}$ for all $k \in \Nat$.
\end{corollary}

\section{Online Schedulers} \label{sec:online}

From this section on we concentrate on online schedulers
that only know the past of the computation. Formally, a scheduler $\sigma$
is {\em online} if for every tree $t$ with
$\sigma(t) = (s_1 \Rightarrow  \ldots \Rightarrow s_k)$
and for every  $1 \leq i <k$, the task
$\sigma(t)[i]$ depends only on $s_1 \Rightarrow  \ldots \Rightarrow s_i$
and on the restriction of the labelling function $L$ to
$\bigcup_{j=1}^i s_j$.

\noindent \emph{Compact Task Systems.}
Any task system can be transformed into a so-called {\em compact} task system such that
for
every scheduler of the compact task system we can construct a scheduler of the original system with
nearly the same properties.
A type $W$ is {\em compact} if there is a rule $X \btran{} \langle Y, Z \rangle$
such that~$X$ is reachable from $W$. A task system is {\em compact} if all its types are compact.
{\em From now on we assume that task systems are compact.}
This assumption is essentially without loss of generality,
 as we argue in \iftechrep{Appendix~\ref{sub:justify-compact}}{\cite{BEKL10:IcalpTechRep}}.

\subsection{Tail Bounds for Online Schedulers} \label{sec:online-bounds}

The following main theorem gives computable lower and upper bounds which hold uniformly for all online
schedulers~$\sigma$.

\begin{theorem} \label{thm:online}
 Let $\Delta$ be subcritical.
 \begin{itemize}
  \item
   Let $\vv, \vw \in (1, \infty)^\Gamma$ be vectors with $\vf(\vv) \le \vv$ and $\vf(\vw) \ge \vw$.
   Denote by $\vvmin$ and $\vwmax$ the least component of~$\vv$ and the greatest component of~$\vw$, respectively.
   Then
   \[
     \frac{\vw_{X_0} - 1}{\vwmax^{k+2} - 1} \le \pr{S^\sigma \ge k}
     \le \frac{\vv_{X_0} - 1}{\vvmin^k - 1} \text{ for all online schedulers~$\sigma$.}
   \]
  \item
   Vectors $\vv, \vw \in (1, \infty)^\Gamma$ with $\vf(\vv) \le \vv$ and $\vf(\vw) \ge \vw$ exist and can be computed in polynomial time.
 \end{itemize}
\end{theorem}
{\em Proof sketch.}
Choose $h > 1$ and $\vu \in (0,\infty)^\Gamma$ such that $h^{\vu_X} = \vv_X$ for all $X \in \Gamma$.
Define for all \mbox{$i \ge 1$} the variable $\ms{i} = \zs{i} \thickdot \vu$ where ``$\mathord{\thickdot}$'' denotes the scalar product,
 i.e., $\ms{i}$ measures the number of tasks at time~$i$
 weighted by types according to~$\vu$.
One can show that $h^{\ms{1}}, h^{\ms{2}}, \ldots$ is a supermartingale for any online scheduler~$\sigma$,
 and, using the Optional Stopping Theorem~\cite{book:Williams}, that $\pr{\sup_i \ms{i} \ge x} \le (\vv_\Init - 1) / (h^x - 1)$ for all~$x$
 (see \iftechrep{the appendix}{\cite{BEKL10:IcalpTechRep}} for the details and \cite{Feller:book,Spitzer:book} for a similar argument on random walks).
As each type has at least weight~$\vumin$, we have that $S^\sigma \ge k$ implies $\sup_i \ms{i} \ge k \vumin$.
Hence 
 $\pr{S^\sigma \ge k} \le \pr{\sup_i \ms{i} \ge k \vumin} \le (\vv_{\Init} - 1)/(\vvmin^k - 1)$.
The lower bound is shown similarly.
\qed

\vspace{1mm}
All online schedulers perform within the bounds of Theorem~\ref{thm:online}. 
For an application of the upper bound, assume one wants to provide as much space as is necessary to guarantee that, say, 99.9\% of the executions of a task system
 can run without needing additional memory.
This can be accomplished, regardless of the scheduler, by providing $k$ space units,
 where $k$ is chosen such that the upper bound of Theorem~\ref{thm:online} is at most~$0.001$.

A comparison of the lower bound with Corollary~\ref{cor:optimal-expectation} proves for subcritical task systems
that the asymptotic performance of any online scheduler $\sigma$ is far away from that of the optimal
offline scheduler: the ratio $\pr{S^\sigma \ge k}/\pr{\xo \ge k}$ is unbounded.

\vspace{-2mm}
\begin{example}
Consider again the task system with pgf $f(x) =  px^2 + q$. For $p < 1/2$ the pgf has two fixed points,
$1$ and $q/p$. In particular, $q/p > 1$, 
so $q/p$ can be used
to obtain both an upper and a lower bound for online schedulers.
Since there is only one type of tasks, vectors have only one component, and
the maximal and minimal components coincide; moreover, in this case the exponent $k+2$ of the
lower bound can be improved to $k$. So the upper and lower bounds coincide, and
we get $\pr{S^\sigma \geq k} = \frac{q/p - 1}{(q/p)^{k}-1}$ for every online scheduler $\sigma$.
In particular, as one intuitively expects, all online schedulers are equivalent.\footnote{For this example $\pr{S^\sigma \geq k}$ can also be computed by elementary means.}
\end{example}

\newcommand{\Gone}{{X_{\mathit min}}}

\subsection{Tail Bounds for Light-First Schedulers} \label{sec:light-first}

We present a class of online schedulers for which a sharper
upper bound than the one given by Theorem~\ref{thm:online} can be proved.
It may be intuitive that a good heuristic is to pick the task with the smallest expected completion time.
If we compute a vector~$\vv$ with $\vf(\vv) \le \vv$ in polynomial time according to the proof of Theorem~\ref{thm:online},
 then the type $\Gone$ for which $\vv_\Gone = \vvmin$ holds turns out to be the type with smallest expected completion time.
This suggests choosing the active type $X$ with smallest component in~$\vv$.
So we look at $\vv$ as a vector of weights, and always choose the lightest active type.
In fact, for this (intuitively good) scheduler we obtain two different upper bounds.

Given a vector~$\vv$ with $\vf(\vv) \le \vv$ we denote by
$\mathord{\sqsubseteq}$ a total order on~$\Gamma$ such that whenever $X \sqsubseteq Y$ then
$\vv_X \le \vv_Y$. If $X \sqsubseteq Y$, then we say that $X$ is lighter than~$Y$.
The {\em $\vv$-light-first scheduler} is an online scheduler that, in each step,
picks a task of the lightest type available in the pool according to~$\vv$.
Theorem~\ref{thm:light-first} below strengthens the upper bound of Theorem~\ref{thm:online}
for light-first schedulers. For the second part of Theorem~\ref{thm:light-first} we use the
notion of {\em $\vv$-accumulating types}. A type~$X \in \Gamma$ is $\vv$-accumulating if for
every $k \geq 0$ the $\vv$-light-first scheduler has a nonzero probability of reaching a
state with at least $k$ tasks of type $X$ in the pool.

\begin{theorem} \label{thm:light-first}
 Let $\Delta$ be subcritical and $\vv \in (1, \infty)^\Gamma$ with $\vf(\vv) \le \vv$.
 Let $\sigma$ be a $\vv$-light-first scheduler.
 Let $\vvminmax := \min_{X \btran{} \langle Y, Z \rangle} \max \{ \vv_Y, \vv_Z \}$
 (here the minimum is taken over all transition rules with two types on the right hand side).
 Then $\vvminmax \ge \vvmin$ and for all~$k \ge 1$
 \[
    \pr{S^\sigma \ge k} \le \frac{\vv_{\Init} - 1}{\vvmin \vvminmax^{k-1} - 1}\,.
 \]

 Moreover, let $\vvminacc := \min \{ \vv_X \mid X \in \Gamma,\ X \text{is $\vv$-accumulating} \}$.
 Then $\vvminacc \ge \vvminmax$, $\vvminacc$ can be computed in polynomial time, and there is an integer~$\ell$ such that for all~$k \ge \ell$
 \[
    \pr{S^\sigma \ge k} \le \frac{\vv_{\Init} - 1}{\vvmin^\ell \vvminacc^{k-\ell} - 1} \,.
 \]
\end{theorem}
{\em Proof sketch.}
 Recall the proof sketch of Theorem~\ref{thm:online} where we used that $S^\sigma \ge k$ implies $\sup_i \ms{i} \ge k \vumin$,
  as each type has at least weight~$\vumin$.
 Let $\ell$ be such that no more than $\ell$ tasks of non-accumulating type can be in the pool at the same time.
 Then $S^\sigma \ge k$ implies $\sup_i \ms{i} \ge \ell \vumin + (k-\ell) \vuminacc$
  which leads to the final inequality of Theorem~\ref{thm:light-first} in a way analogous to the proof sketch of Theorem~\ref{thm:online}.
\qed

Intuitively, a light-first scheduler ``works against'' light tasks by picking them as soon as possible.
In this way it may be able to avoid the accumulation of some light types, so it may achieve $\vvminacc > \vvmin$.
This is illustrated in the following example.

\vspace{-2mm}
\begin{example}
Consider the task system with 2~task types and pgfs
$x  =  a_2 xy + a_1 y + a_0$ and $y  =  b_2 xy + b_1 y + b_0$, where
$a_2+a_1+a_0 = 1=b_2+b_1+b_0 = 1$.
The system is subcritical if $a_1b_2 < a_2b_1 - a_2 + b_0$. The pgfs have a
greatest fixed point~$\vv$ with
$\vv_X =   (1 - a_2 - b_1 - a_1b_2 + a_2b_1)/b_2$ and
$\vv_Y = (1-b_1-b_2)/(a_2 + a_1b_2 - a_2b_1)$.
We have $\vv_X \le \vv_Y$ iff
$a_2 - b_2 \le a_2b_1 - a_1b_2$, and so the light-first scheduler chooses $X$ before $Y$ if
this condition holds, and $Y$ before $X$ otherwise. We show that the light-first scheduler is
asymptotically optimal. Assume w.l.o.g.\ $\vv_X \le \vv_Y$.
Then $X$ is not accumulating (because $X$-tasks are picked as soon as they are created), and so $\vvminacc = \vv_Y$.
So the upper bound for the light-weight scheduler yields a constant~$c_2$ such that $\pr{S^\sigma \ge k} \le c_2 / {\vv_Y^{k}}$.
But the general lower bound for arbitrary online schedulers states that there is a constant~$c_1$
 such that $\pr{S^\sigma \ge k} \ge c_1 / \vv_Y^k$, so we are done.
\end{example}

\subsection{Tail Bounds for Depth-first Schedulers} \label{sub:depth-first}

Space-efficient scheduling of multithreaded computations
has received considerable attention \cite{NarBel99,BluLei99,AroBluPla02,ALHH08}. The setting of
these papers is slightly different from ours, because they assume
data dependencies among the threads, which may cause a thread to wait
for a result from another thread. In this sense our setting is similar to
that of \cite{KarpZhang93}, where, in thread terminology, the threads can execute
independently.

These papers focus on {\em depth-first} computations, in which
if thread~$A$ has to wait for thread~$B$, then $B$ was spawned by~$A$ or
by a descendant of~$A$. The
optimal scheduler is the one that, when
$A$ spawns~$B$, interrupts the execution of~$A$ and continues with~$B$;
this
online scheduler produces the familiar stack-based execution
\cite{BluLei99,NarBel99}.

We study the performance of this {\em depth-first} scheduler.
Formally, a depth-first scheduler $\sigma_\lambda$ is determined by
a function~$\lambda$ that assigns to each rule
$r= X \btran{} \langle Y,Z\rangle$ either $YZ$ or $Z\,Y$.
If $\lambda(r)=YZ$, then $Z$ models the continuation of the thread $X$,
while $Y$ models
a new thread for whose termination $Z$ waits.
%
The depth-first scheduler $\sigma_\lambda$ keeps as an internal data
structure
a word $w \in \Gamma^*$, a ``stack'', such that the Parikh image of~$w$ is
the multiset of
the task types in the pool.
If $w = Xw'$ for some $w' \in \Gamma^*$, then $\sigma$ picks~$X$.
If a transition rule $X \btran{} \alpha$ ``fires'', then $\sigma_\lambda$
replaces
$Xw'$ by $\beta w'$ where $\beta=\lambda(X \btran{} \alpha)$.

\noindent Using techniques of~\cite{BEK09:fsttcs} for {\em probabilistic pushdown
systems},
we obtain the following:
\newcommand{\stmtthmdepthfirst}{
  Let $\Delta$ be subcritical and $\sigma$ be any depth-first scheduler.
  Then $\pr{S^\sigma = k}$ can be computed in time $\bigo(k \cdot |\Gamma|^3)$
  in the unit-cost model.
  Moreover, there is $0 < \rho < 1$ such that
   $\pr{S^\sigma \ge k} \in \Theta(\rho^k)$,
  i.e, there are $c,C > 0$ such that
   $c \rho^k \le \pr{S^\sigma \ge k} \le C \rho^k$ for all~$k$.
  Furthermore, $\rho$ is the spectral radius of a nonnegative matrix~$B \in
\R^{\Gamma \times \Gamma}$, where $B$ can be computed in polynomial time.
}
\begin{theorem}\label{thm:depth-first}
  \stmtthmdepthfirst
\end{theorem}
While the proof of Theorem~\ref{thm:depth-first} does not conceptually
 require much more than the results of~\cite{BEK09:fsttcs},
 the technical details are delicate.
The proof can be found in \iftechrep{the appendix}{\cite{BEKL10:IcalpTechRep}}.

\section{Expectations} \label{sec:expectations}
In this section we study the expected completion space, i.e.,
the expectation~$\Ex{S^{\sigma}}$ for  both offline and online
schedulers.  Fix a task system $\Delta=(\Gamma,\btran{},\Prob, \Init)$.

\emph{Optimal (Offline) Schedulers.}
The results of Section~\ref{sec:optimal} allow to efficiently approximate
the expectation~$\Ex{\xo}$. Recall that for any random variable~$R$ with values
in the natural numbers we have $\Ex{R} = \sum_{i=1}^\infty \pr{R \ge i}$.
So we can (under-) approximate $\Ex{R}$ by $\sum_{i=1}^k \pr{R \ge i}$ for
finite~$k$. We say that {\em $k$ terms compute $b$ bits of $\Ex{\xo}$} if
$\Ex{\xo} - \sum_{i=0}^{k-1} (1 - \ns{i}_{\Init}) \le 2^{-b}$.

\begin{theorem} \label{thm:whitebox-exp}
The expectation $\Ex{\xo}$ is finite (no matter whether $\Delta$ is
critical or subcritical). Moreover, $\bigo(b)$ terms compute $b$ bits of
$\Ex{\xo}$. If the task system $\Delta$ is subcritical, then $\log_2 b + \bigo(1)$
terms compute $b$ bits of $\Ex{\xo}$.
Finally, computing $k$ terms takes time $\bigo(k \cdot |\Gamma|^3)$ in the unit cost model.
\end{theorem}


\emph{Online Schedulers.} The main result for online schedulers states that
the finiteness of $\Ex{S^{\sigma}}$ does not depend on the choice of the online
scheduler $\sigma$.
%
\begin{theorem} \label{thm:online-equivalence}
If $\Delta$ is subcritical, then $\Ex{S^{\sigma}}$ is finite for every online scheduler~$\sigma$.
If $\Delta$ is critical, then $\Ex{S^{\sigma}}$ is infinite for every online
scheduler~$\sigma$.
\end{theorem}
\noindent {\em Proof sketch.}
 The first assertion follows from Theorem~\ref{thm:online}.
 Let $\Delta$ be critical.
 For this sketch we focus on the case where $X_0$ is reachable from every type.
 By Proposition~\ref{prop:critical-spectral} the spectral radius of $\vf'(\vone)$ equals~$1$.
 Then Perron-Frobenius theory guarantees the existence of a vector~$\vu$ with $\vf'(\vone) \vu = \vu$ and $\vu_X > 0$ for all~$X$.
 Using a martingale argument, similar to the one of Theorem~\ref{thm:online}, one can show that
  the sequence $\ms{1}, \ms{2}, \ldots$ with $\ms{i} := \zs{i} \thickdot \vu$ is a martingale for every scheduler~$\sigma$,
 and, using the Optional-Stopping Theorem, that $\pr{S^\sigma \ge k} \ge \vu_{X_0} / (k+2)$.
 So we have $\Ex{S^{\sigma}} = \sum_{k=1}^\infty \pr{S^\sigma \ge k} \ge \sum_{k=1}^\infty \vu_{X_0}/(k+2) = \infty$.
\qed
Since we can decide
in polynomial time whether a system is subcritical or critical, we can do the same
to decide on the finiteness of the expected completion time.

\emph{Depth-first Schedulers.}
To approximate $\Ex{S^{\sigma}}$ for a given depth-first scheduler~$\sigma$,
 we can employ the same technique as for optimal offline schedulers, i.e.,
 we approximate $\Ex{S^\sigma}$ by $\sum_{i=1}^k \pr{S^\sigma \ge i}$ for finite~$k$.
We say that {\em $k$ terms compute $b$ bits of $\Ex{S^{\sigma}}$} if
 $\Ex{S^{\sigma}} - \sum_{i=1}^k \pr{S^\sigma \ge i} \le 2^{-b}$.
%
%
\begin{theorem}[see Theorem~19 of~\cite{BEK09:fsttcs}] \label{thm:depth-first-expectation}
 Let $\Delta$ be subcritical, and let $\sigma$ be a depth-first scheduler.
 Then $\bigo(b)$ terms compute $b$ bits of~$\Ex{S^\sigma}$,
  and computing $k$ terms takes time $\bigo(k \cdot |\Gamma|^3)$ in the unit cost model.
\end{theorem}

\section{Conclusions} \label{sec:conclusions}

We have initiated the study of scheduling tasks that can stochastically
generate other tasks. We have provided strong results on the performance
of both online and offline schedulers for the case of one processor and
task systems with completion probability~1.
It is an open problem how to compute and analyze online schedulers which are optimal in a sense. While we profited from
the theory of branching processes, the theory considers (in computer
science terms) systems with an unbounded number of processors, and
therefore many questions had not been addressed before or even posed.

\vspace{3mm}
\noindent {\em Acknowledgement.} We thank the referees for their helpful comments.

\bibliographystyle{plain} 
\bibliography{db}
\iftechrep{
 \newpage
 \appendix
 \section{Proofs of Section~\ref{sec:prelim}} \label{sec:app-prelim}

\subsection{Proof of Proposition~\ref{prop:critical-spectral}}
\begin{qproposition}[ (\cite{Harris63,EYstacs05Extended})]{\ref{prop:critical-spectral}}
Let $\Delta$ be a task system with pgf~$\vf$.
Denote by $\vf'(\vone)$ the Jacobian matrix of partial derivatives of~$\vf$ evaluated at~$\vone$.
If $\Delta$ is critical, then the spectral radius of $\vf'(\vone)$ is equal to~$1$;
otherwise it is strictly less than~$1$.
It can be decided in polynomial time whether $\Delta$ is critical.
\end{qproposition}
\begin{proof}
One can show (see e.g.~\cite{EKM04}) that $\Ex{T_X}$ is the $X$-component of the least nonnegative fixed point of \mbox{$\vf'(\vone) \vx + \vone$}, i.e.,
  the $X$-component of the (componentwise) least vector $\vx \in [0, \infty]^\Gamma$ with $\vx = \vf'(\vone) \vx + \vone$.
This least fixed point is given by $\sum_{i=0}^\infty (\vf'(\vone))^i \vone$, a series that may or may not converge.
It is a standard fact (see e.g.~\cite{book:HornJ}) that the series converges iff
 $\rho(\vf'(\vone)) < 1$ holds for the spectral radius $\rho(\vf'(\vone))$ of~$\vf'(\vone)$.

Assume first that $\Delta$ is subcritical.
Then the above series must converge, so we have $\rho(\vf'(\vone)) < 1$ in this case.
Now assume that $\Delta$ is critical.
Then the above series must diverge, so we have $\rho(\vf'(\vone)) \ge 1$.
On the other hand, in \cite{EKL08:stacs,EYstacs05Extended} it is shown that $\rho(\vf'(\vone)) \le 1$.
(More precisely, it is shown there that $\rho(\vf'(\vy)) < 1$ holds for $\vy$ that are strictly less than the least fixed point of $\vf$.
 By continuity of eigenvalues, $\rho(\vf'(\vy)) \le 1$ also holds for the least fixed point of $\vf$ which is $\vone$ according to
 the proof of Proposition~\ref{prop:prelim-assumptions}.)
Hence we have $\rho(\vf'(\vone)) = 1$.

In order to decide on the criticality, it thus suffices to decide whether the spectral radius of~$\vf'(\vone)$ is $\ge 1$.
This condition holds iff $\vf'(\vone) \vx \ge \vx$ holds for a nonnegative, nonzero vector~$\vx$
(see e.g.\ Thm.~2.1.11 of~\cite{book:BermanP} and cf.~\cite{EYstacs05Extended}).
This can be checked in polynomial time with linear programming.
\qed
\end{proof}

 \section{Proofs of Section~\ref{sec:optimal}} \label{sec:app-optimal}

\subsection{Proof of Proposition~\ref{prop:char-optimal}}

\begin{qproposition}{\ref{prop:char-optimal}}
 Let $t$ be a family tree.
 Then
 \[
  \xo(t) =
   \begin{cases}
    \min\left\{ \begin{array}{l}
                  \max\{\xo(t_0)+1, \xo(t_1)\}, \\
                  \max\{\xo(t_0), \xo(t_1)+1\}
                \end{array}
        \right\}
             & \text{if $t$ has two children $t_0$, $t_1$} \\
    \xo(t_0) & \text{if $t$ has exactly one child $t_0$} \\
    1        & \text{if $t$ has no children.}
   \end{cases}
 \]
\end{qproposition}
\begin{proof}
 Recall the proof sketch from the main body of the paper.
 We detail the argument why one of the two given scheduling strategies is optimal,
  i.e., we argue why the scheduler cannot save space by interleaving the schedulings for $t_0$ and~$t_1$.

 Consider an optimal scheduling of~$t$.
 W.l.o.g.\ the task $t_0$ terminates first.
 Then at least one $t_1$-task sticks around during the whole derivation of~$t_0$.
 So this scheduling needs space of at least $\xo(t_0)+1$.
 Obviously, any scheduling of~$t$ needs space of at least~$\xo(t_1)$.
 So the optimal scheduler needs space of at least $\max\{\xo(t_0)+1, \xo(t_1)\}$.
 But this lower bound is matched by the scheduling strategy given in the main body of the paper.
\qed
\end{proof}

\subsection{Proof of Theorem~\ref{thm:whitebox}}
\begin{qtheorem}{\ref{thm:whitebox}}
 $\pr{\xo_X \leq k} = \ns{k}_X$ for every type $X$ and every $k \geq 0$.
\end{qtheorem}
\begin{proof}
\renewcommand{\L}{\ell}
Let us inductively define the function $\L$ on trees as follows.
\[
 \L(t) :=
  \begin{cases}
    0           & \text{ if $t$ has no children } \\
    \L(t_0) + 1 & \text{ if $t$ has one child } \\
    \L(t_0) + 1 & \text{ if $t$ has two children and } \xo(t_0) > \xo(t_1) \\
    \L(t_1) + 1 & \text{ if $t$ has two children and } \xo(t_0) < \xo(t_1) \\
    0           & \text{ if $t$ has two children and } \xo(t_0) = \xo(t_1) \,.
  \end{cases}
\]
With Proposition~\ref{prop:char-optimal}, $\L(t)$ is the length of
a longest path from the root to a descendant with the same
$\xo$-value.

We proceed by induction on $k$. The base case $k=0$ is trivial.
Let $k \ge 0$ and let $t$ be an $X$-tree with $\xo(t) = k+1$.
We have to show $\pr{\xo_X = k+1} = \Ds{k+1}_X$ where
 \[
  \Ds{k+1} = \sum_{i=0}^\infty \vf'(\ns{k})^i \left(\vf(\ns{k}) - \ns{k}\right)\,.
 \]
We show the following stronger claim:
 \[
  \pr{\xo_X(t) = k+1,\ \L(t) = i} = \left( \vf'(\ns{k})^i \left(\vf(\ns{k}) - \ns{k}\right) \right)_X\,.
 \]
We proceed by an (inner) induction on $i$. For the induction base
$i=0$ we first dispense with the case $k=0$. We have
 \[
   \pr{\xo_X(t) = 1,\ \L(t) = 0}  = \pr{\text{$t$ has no children}}
 \]
 because if $t$ has one child then $\L(t) \ge 1$ and if $t$ has two children, then $\xo_X(t) \ge 2$.
With the definition of~$\vf$ we obtain
 \[
   \pr{\xo_X(t) = 1,\ \L(t) = 0} = \sum_{X\btran{p}\epsilon} p = \vf_X(\vzero) = \vf_X(\ns{0}) - \ns{0}_X \,.
 \]
Now we complete the induction base $i=0$ with the case $k \ge 1$.
We have
 \begin{equation} \label{eq:proof-optimal-hyp3}
   \pr{\xo_X(t) = k+1,\ \L(t) = 0} = \pr{\text{$t$ has two children},\ \xo(t_0) = \xo(t_1) = k}
 \end{equation}
 because if $t$ has one child, then $\L(t) \ge 1$, and if $t$ has no children, then $\xo_X(t) = 1$.
Further we have by Proposition~\ref{prop:char-optimal}
 \begin{eqnarray} \label{eq:proof-optimal-hyp4}
   \pr{\xo_X(t) \le k} & = & \sum_{X\btran{p} \langle Y, Z \rangle} p \cdot \bigl( \pr{\xo_Y(t_0) \le k} \pr{\xo_Z(t_1) \le k} \nonumber\\
                       && \hspace{23mm} - \pr{\xo_Y(t_0) = k} \pr{\xo_Z(t_1) = k} \bigr) \\
                       && \quad + \sum_{X\btran{p}Y} p \cdot \pr{\xo_Y(t_0) \le k} \nonumber \\
                       && \quad + \sum_{X\btran{p}\emptyset} p \nonumber \,.
 \end{eqnarray}
Combining these equations we obtain
 \begin{align*}
  \pr{\xo_X(t) = k+1,\ \L(t) = 0} & = \sum_{X\btran{p}\langle Y, Z \rangle} p \cdot \pr{\xo_Y(t_0) = k} \pr{\xo_Z(t_1) = k}
                                                 && \text{(by \eqref{eq:proof-optimal-hyp3})} \\
                                  & = \sum_{X\btran{p} \langle Y, Z \rangle} p \cdot \pr{\xo_Y(t_0) \le k} \pr{\xo_Z(t_1) \le k}
                                                 && \text{(by \eqref{eq:proof-optimal-hyp4})} \\
                                  & \quad + \sum_{X\btran{p}Y} p \cdot \pr{\xo_Y(t_0) \le k}  + \sum_{X\btran{p}\epsilon} p \\
                                  & \quad - \pr{\xo_X(t) \le k} \\
                                  & = \sum_{X\btran{p} \langle Y, Z \rangle} p \cdot \ns{k}_Y \ns{k}_Z && \text{(ind.~hyp.~on~$k$)} \\
                                  & \quad + \sum_{X\btran{p}Y} p \cdot \ns{k}_Y  + \sum_{X\btran{p}\epsilon} p \\
                                  & \quad - \ns{k}_X \\
                                  & = \vf_X(\ns{k}) - \ns{k}_X && \text{(def.~of~$\vf$)}
 \end{align*}
For the induction step, let $i \ge 0$. Then by
Proposition~\ref{prop:char-optimal} and the definition of~$\L$
 \begin{align*}
  \pr{\xo_X(t) = k+1,\ \L(t) = i+1} \hspace{-55mm} \\
                                    & = \sum_{X\btran{p} \langle Y, Z \rangle} p \cdot \left( \pr{\xo_Y(t_0) \le k} \pr{\xo_Z(t_1) = k+1,\ \L(t_1) = i} \right.\\
                                    & \hspace{23mm}                 \left. {} + \pr{\xo_Y(t_0) = k+1,\ \L(t_0) = i} \pr{\xo_Z(t_1) \le k} \right) \\
                                    & \quad + \sum_{X\btran{p}Y} p \cdot \pr{\xo_Y(t_0) = k+1,\ \L(t_0) = i}  \\
                                    & = \sum_{X\btran{p} \langle Y, Z \rangle} p \cdot
                                           \left( \ns{k}_Y \left( \vf'(\ns{k})^i \left(\vf(\ns{k}) - \ns{k}\right) \right)_Z \right. \\
                                    & \hspace{23mm}                 \left. {} + \left( \vf'(\ns{k})^i \left(\vf(\ns{k}) - \ns{k}\right) \right)_Y \ns{k}_Z \right)
                                                                                                                                && \text{(ind.~hyp.~on~$k$, $i$)} \\
                                    & \quad + \sum_{X\btran{p}Y} p \cdot \left( \vf'(\ns{k})^i \left(\vf(\ns{k}) - \ns{k}\right) \right)_Y \\
                                    & = \sum_{Y\in\Gamma} \vf'_{XY}(\ns{k}) \left( \vf'(\ns{k})^i \left(\vf(\ns{k}) - \ns{k}\right) \right)_Y
                                                                                                                                && \text{(def.~of~$\vf$)} \\
                                    & = \vf'_X(\ns{k})  \vf'(\ns{k})^i \left(\vf(\ns{k}) - \ns{k}\right) \\
                                    & = \left( \vf'(\ns{k})^{i+1} \left(\vf(\ns{k}) - \ns{k}\right) \right)_X \,.
 \end{align*}
\qed
\end{proof}

\subsection{Proof of Corollary~\ref{cor:optimal-expectation}}
\begin{qcorollary}{\ref{cor:optimal-expectation}}
 For any task system~$\Delta$ there are real numbers $c > 0$ and $0 < d < 1$ such that $\pr{\xo_X \ge k} \le c \cdot d^k$ for all $k \in \Nat$.
 If $\Delta$ is subcritical, then there are real numbers $c > 0$ and $0 < d < 1$ such that $\pr{\xo_X \ge k} \le c \cdot d^{2^k}$ for all $k \in \Nat$.
\end{qcorollary}
\begin{proof}
 By Theorem~\ref{thm:whitebox} we have $\pr{\xo \ge k} = 1 - \ns{k-1}_\Init \le 1 - \ns{k}_\Init$.
 So the corollary can be understood as a statement
  on the convergence speed of Newton's method for solving $\vx = \vf(\vx)$.
 The fact that Newton's method started at $\vzero$ converges to~$\vone$ (the least fixed point of~$\vf$)
  is shown in~\cite{EYstacs05Extended}.

 For the subcritical case, observe that the matrix $\Id - \vf'(\vone)$ is nonsingular
  because otherwise $1$ would be an eigenvalue of $\vf'(\vone)$
  which would, together with Proposition~\ref{prop:critical-spectral},
  contradict the assumption that the task system is subcritical.
 For nonsingular systems, it is a standard fact (see e.g.~\cite{OrtegaRheinboldt:book})
  that Newton's method converges quadratically.
 As $\pr{\xo \ge k} \le 1 - \ns{k}_\Init$, the statement follows.

 For the general case (subcritical or critical) Newton's method for solving $\vx = \vf(\vx)$ has been
  extensively studied in~\cite{KLE07:stoc,EKL08:stacs} and it follows from there that
  there is a $c_1 \in (0,\infty)$ such that $1 - \ns{k}_X \le c_1 \cdot 2^{-k/(n2^n)}$ where $n = |\Gamma|$,
  implying the statement.
\end{proof}

 \section{Proofs of Section~\ref{sec:online}} \label{sec:app-online}

\subsection{A Characterization of Online Schedulers}

For proofs involving online schedulers~$\sigma$,
 it is convenient to work with a function $\fsch_\sigma$ (defined below)
 which essentially characterizes~$\sigma$.
To define it, fix an online scheduler~$\sigma$. For every tree $t$ with
$\sigma(t) = (s_1 \Rightarrow  \ldots \Rightarrow s_k)$ and for every
$j \geq 0$, let $\zs{j}(t)$ denote the multiset of types
labelling the tasks of $s_j$ if $j \leq k$ (i.e., $\zs{j}(t) = \langle L(w) \mid w \in s_j \rangle$), and the empty multiset
otherwise. One can show that an online scheduler $\sigma$ induces a partial function
$\fsch_\sigma \colon (\N^\Gamma)^* \rightarrow \Gamma$ defined as
follows:  $\fsch_\sigma(\css{1} \ldots \css{i})$ is defined if
there is a tree $t$ such that
$\sigma(t) = (s_1 \Rightarrow  \ldots \Rightarrow s_k)$ with $k\geq i$ and
$\css{1} = \zs{1}(t), \ldots, \css{i} = \zs{i}(t)$; in this case
$\fsch_\sigma(\css{1} \ldots \css{i}) = L(\sigma(t)[i])$.
Intuitively, if $\fsch_\sigma$ gets as input the
multisets of types of the states $s_1, \ldots, s_i$, then it returns the type
of the task of $s_i$ picked up by the scheduler.
Let $\Xs{i} = \fsch_\sigma(\zs{1}, \ldots, \zs{i})$, i.e., $\Xs{i}$ is the type picked up at the $i$-th step.
Then $\Xs{i}$ is randomly replaced by new types according to the distribution on the transition rules.
More precisely, if $\rs{i}:=\zs{i+1} + \Xs{i} - \zs{i}$, then
$\pr{\rs{i}=\alpha \mid \Xs{i}=X} = \sum_{X \btran{p} \alpha} p$.

We will show the following proposition, which allows us to identify an online
scheduler~$\sigma$ with the function~$\fsch_\sigma$.

\begin{proposition}\label{prop:f_sigma}
Let $\sigma_1, \sigma_2$ be online schedulers.
If $\fsch_{\sigma_1} = \fsch_{\sigma_2}$, then $\pr{S^{\sigma_1}=k}  = \pr{S^{\sigma_2}=k}$ for all $k \geq 1$.
\end{proposition}

%
\begin{lemma}\label{lem:uniqueness}
Let $\sigma$ be an online scheduler. For every family tree $t$ the
first $i\geq 1$ states of $\sigma(t)$ are uniquely determined by
$\zs{1}(t),\ldots,\zs{i}(t)$. In particular, the function $\fsch_{\sigma}$ is
well-defined.
\end{lemma}
\begin{proof}
We proceed by induction on $i$. The case $i=1$ is trivial.
Let us consider $\zs{1}(t),\ldots,\zs{i+1}(t)$, and
let $d=(s_1\Rightarrow \cdots \Rightarrow s_i\Rightarrow s_{i+1})$ be a prefix
of the derivation $\sigma(t)$. By induction,
$s_1\Rightarrow \cdots \Rightarrow s_i$ is completely determined by
$\zs{1}(t),\ldots,\zs{i}(t)$.
By the definition of online scheduler,
$\sigma(t)[i]$ is completely determined by
$s_1\Rightarrow \cdots \Rightarrow s_i$ and $\zs{1}(t),\ldots,\zs{i}(t)$.
Finally, there is a unique transition rule
$L(\sigma(t)[i])\btran{} \alpha$ where $\alpha=\zs{i+1}(t)-\zs{i}(t)+
\langle L(\sigma(t)[i])\rangle$. But then $s_{i+1}$ is also
uniquely determined.
\qed
\end{proof}
\begin{lemma}\label{lem:prob_cylinder}
Let $\css{1}\cdots\css{i}\in (\N^{\Gamma})^+$
such that for every $1\leq j<i$
the value $\fsch_{\sigma}(\css{1}\cdots \css{j})$ is defined. Then
$\pr{\bigwedge_{j=1}^{i} \zs{j}=\css{j}}=\prod_{j=1}^{i-1}
\Prob(\fsch_{\sigma}(\css{1}\cdots \css{j})\btran{} \alpha_{j})$
where for every $1\leq j<i$ we have
$\alpha_{j}=\css{j+1}-\css{j}+\langle
\fsch_{\sigma}(\css{1}\cdots \css{j})\rangle$.
\end{lemma}
\begin{proof}
Let us denote by $\mathcal{R}$ the set of all family
trees $t$ such that $\zs{j}(t)=\css{j}$ for $1\leq j\leq i$.
By Lemma~\ref{lem:uniqueness}, there is a derivation
$d=s_1\Rightarrow \cdots \Rightarrow s_i$ and a function
$l:\bigcup_{j=1}^i s_j\rightarrow \Gamma$ such that
for every $t=(N,L)\in \mathcal{R}$ we have that $d$
is a prefix of $\sigma(t)$ and $l$ coincides with $l$ on
the subtree $\bigcup_{j=1}^i s_j$. Let us denote by
$t^s$ the tree $\bigcup_{j=1}^i s_j$. Note that $t^s$
is a subtree of every tree of $\mathcal{R}$ rooted in $\epsilon$.
Let us denote by $\mathcal{I}$ the set of all inner nodes of $t^s$.
For every $v\in \mathcal{I}$, we denote by ${child}(v):=
\langle l(va)\mid a\in \{0,1\},va\in t^s\rangle$ the multiset of
labels of children of the node $v$ in $t^s$. Let us denote by $\mathcal{L}$
the set of all leaves of $t^s$. It follows directly
from the definition of $\Pr$, that for all $t\in \mathcal{R}$
we have
\[
\pr{t}=\prod_{v\in \mathcal{I}} \Prob(L(v)\btran{} {child}(v))\cdot
\prod_{v\in \mathcal{L}} \pr{t_v}
\]
However, it follows directly from definitions that for every
$v\in \mathcal{I}$ there is precisely one $1\leq j<i$ such that
$\sigma(t)[j]=v$, and then $L(v)=\fsch_{\sigma}(\css{1}\cdots \css{j})$
and ${child}(v)=\alpha_j$. Therefore,
\[
\pr{t}=\prod_{j=1}^{i-1}
\Prob(\fsch_{\sigma}(\css{1}\cdots \css{j})\btran{} \alpha_{j})\cdot
\prod_{v\in \mathcal{L}} \pr{t_v}
\]
Finally,
\[
\sum_{t\in \mathcal{R}} \pr{t}=
\prod_{j=1}^{i-1}
\Prob(\fsch_{\sigma}(\css{1}\cdots \css{j})\btran{} \alpha_{j})\cdot
\prod_{v\in \mathcal{L}}\ \sum_{t'\in \mathcal{T}_{L(v)}} \pr{t'}=
\prod_{j=1}^{i-1}
\Prob(\fsch_{\sigma}(\css{1}\cdots \css{j})\btran{} \alpha_{j})
\]
\qed
\end{proof}

\noindent Now we can prove Proposition~\ref{prop:f_sigma}.
\begin{proof}[of Proposition~\ref{prop:f_sigma}]
We denote by $\zs{i}_{\lambda}$  the variable $\zs{i}$ evaluated
with respect to a given scheduler $\lambda$.
Let us denote by $A_{def}$ the set of all
$\css{1}\cdots \css{i}\in (\N^{\Gamma})^+$
such that $\fsch_{\sigma_1}(\css{1}\cdots \css{j})=
\fsch_{\sigma_2}(\css{1}\cdots \css{j})$ is defined
for all $1\leq j\leq i-1$, and
$\css{i}=\vzero$.
%
By Lemma~\ref{lem:prob_cylinder}, for every
$\css{1}\cdots \css{i}\in A_{def}$ we have
\begin{align*}
\pr{\bigwedge_{j=1}^{i} \zs{j}_{\sigma_1}=\css{j}}
      & = \prod_{j=1}^{i-1} \Prob(\fsch_{\sigma_1}(\css{1}\cdots \css{j})\btran{}
          \alpha_{j}) \\
      & = \prod_{j=1}^{i-1} \Prob(\fsch_{\sigma_2}(\css{1}\cdots \css{j})\btran{}
          \alpha_{j})   \\
      & = \pr{\bigwedge_{j=1}^{i} \zs{i}_{\sigma_2}=\css{j}}
\end{align*}
%
However, then $\pr{S^{\sigma_1}=k}=\pr{S^{\sigma_2}=k}$ because
the values of $S^{\sigma_1}$ and $S^{\sigma_2}$ are
determined by the values of $\zs{1}_{\sigma_1},
\zs{2}_{\sigma_1},\ldots$ and $\zs{1}_{\sigma_2},
\zs{2}_{\sigma_2},\ldots$, and for 
all family trees $t$ we have that
a prefix of $\zs{1}_{\sigma_1}(t),\zs{2}_{\sigma_1}(t),\ldots$ and
a prefix of $\zs{1}_{\sigma_2}(t),\zs{2}_{\sigma_2}(t),\ldots$ are
in $A_{def}$.
\qed
\end{proof}


\subsection{Justification for Compactness} \label{sub:justify-compact}

In Section~\ref{sec:online} we claimed that we can focus on compact task systems essentially without loss of generality.
We justify this claim now.

A non-compact task system can be compacted by iteratively
removing all rules with non-compact types on the left hand side, and
all occurrences of non-compact types on the right hand side.

%

\begin{proposition} \label{prop:normal-form}
 Let us denote by $\Gamma'$ the set of all task types removed from~$\Delta$
  by the above compacting procedure and let $|\Gamma'| = \ell$.
 If $\Init \in \Gamma'$, then there is a scheduler~$\sigma$ such that $S^\sigma \le \ell$.

 Assume that $\Init\not\in \Gamma'$.
 Let $\Delta'$ be the compacted version of~$\Delta$ (i.e.,
 $\Gamma\setminus\Gamma'$ is
 the set of task types of $\Delta'$).
 Every scheduler $\sigma'$ for~$\Delta'$ can be transformed into a scheduler~$\sigma$ for~$\Delta$ such that for all~$k$
 \[
  \pr{S^{\sigma',\Delta'} \ge k} \le \pr{S^{\sigma,\Delta} \ge k} \le
  \pr{S^{\sigma',\Delta'} \ge k - \ell} \,.
 \]
 (The second superscript of $S$ indicates the task system on which the
 scheduler operates.)
\end{proposition}

Notice that computing $\sigma$ from~$\sigma'$ is easy: $\sigma$ acts like~$\sigma'$ but gives preferences
to the types that have been (first) eliminated during the compacting procedure.

Now we prove Proposition~\ref{prop:normal-form}.

\begin{proof}
 \newcommand{\Gnon}{\Gamma_{\it non}}
 Let $\Delta_1$ be a non-compact task system with a non-compact types~$\Gnon$,
  and let $\Delta_0$ be the (possibly non-compact) task system obtained from~$\Delta_1$
  by removing all rules with non-compact types on the left hand side
  and all occurrences of non-compact types on the right hand side of all rules,
  i.e., $\Delta_0$ is obtained from~$\Delta_1$ by performing the first iteration of the compacting procedure.
 Let $\sigma_0$ be a scheduler for~$\Delta_0$.
 Construct a scheduler~$\sigma_1$ for~$\Delta_1$ as follows:
  \begin{quote}
   The scheduler~$\sigma_1$ acts exactly like~$\sigma_0$ until one or two $\Gnon$-tasks are created
    at which point the completion space of the derivation may be increased by at most~$1$.
   Then $\sigma_1$ picks a $\Gnon$-task, say $\tau_1$.
   Since the $\Gnon$-types are non-compact, $\sigma_1$ can complete $\tau_1$ without further increasing the completion space.
   After $\tau_1$ has been finished, there may be another $\Gnon$-task left, say $\tau_2$, that was
    created at the time when $\tau_1$ was created.
   If there is such a $\tau_2$, then $\sigma_1$ completes $\tau_2$ in the same way it has completed $\tau_1$.
   After $\tau_1$ (and possibly $\tau_2$) have been completed, $\sigma_1$ resumes to act like~$\sigma_0$.
  \end{quote}
 It follows from this construction that the incorporation of the non-compact type~$\Gnon$
  increases the completion space of a derivation by at most~$1$.

 A straightforward induction on this construction shows for the statement of the proposition:
  \[
   \pr{S_X^{\sigma',\Delta'} \le k} \le \pr{S_X^{\sigma,\Delta} \le k + \ell} \text{ for all $X \in \Gamma \setminus \Gamma'$.}
  \]

 If $X_0 \in \Gamma'$, then the above construction also works.
  (It extends a scheduler operating on a possibly empty task system, but this poses no problems.)
 So, again by induction, we obtain a scheduler~$\sigma$ for~$\Delta$ with $S_X^{\sigma,\Delta} \le \ell$ for all $X \in \Gamma'$.

 It remains to show the inequality $\pr{S_X^{\sigma',\Delta'} \ge k} \le \pr{S_X^{\sigma,\Delta} \ge k}$,
  but this is clear because $\Delta'$ is obtained from deleting rules and types from~$\Delta$ and $\sigma$ is
   obtained by extending~$\sigma'$.
\qed
\end{proof}

\subsection{Proof of Theorem~\ref{thm:online}}

We split the proof in several lemmata.
With regard to the computation of a suitable vector~$\vv$ we first prove the following lemma.

\begin{lemma} \label{lem:finding-a-v}
 Let $\vu \in [1,\infty)^\Gamma$ denote the vector of expected completion times, i.e., $\vu_Y = \Ex{T_Y}$ for all $Y \in \Gamma$.
 Then $\vu$ exists and is the unique solution of $\vx = \vf'(\vone)\vx + \vone$.
 Let $Q(\vu,\vu)$ denote the ``quadratic part'' of~$\vf(\vu)$,
  i.e., $\left( Q(\vu,\vu) \right)_X = \sum_{X \btran{p} YZ} p \cdot \vu_Y \cdot \vu_Z$ for all $X,Y,Z \in \Gamma$.
 Let $s := 1 / q_{\mathit max} > 0$ where $q_{\mathit max}$ is the largest component of~$Q(\vu,\vu)$.
 Then for all $r \ge 0$ we have $\vf(\vone + r \vu) \le \vone + r \vu$ iff $r \le s$.
\end{lemma}

Using this lemma a suitable $\vv$ can be found as follows:
First compute~$\vu$ by solving $\vx = \vf'(\vone) \vx + \vone$.
This yields $Q(\vu,\vu)$, and, consequently,~$s$.
With regard to the upper bound of the theorem we are interested in a~$\vv$ which is as large as possible,
 so pick $\vv := \vone + s \vu$. 
All steps can be performed in polynomial time.

\vspace{2mm}
\noindent{\em Proof of the lemma.}
 The fact that $\vu = \vf'(\vone) \vu + \vone$ exists and is the vector of expected completion times
  follows from the remarks made at the beginning of the proof of Proposition~\ref{prop:critical-spectral}.
 Recall that the pgf~$\vf$ is a vector of polynomials
  of degree~2 with positive coefficients.
 So it can be written as
  \[\vf(\vx) = Q(\vx,\vx) + L \vx + \vc \]
 where $Q(\vx,\vx)$ is the quadratic part of~$\vf(\vx)$.
 A straightforward calculation shows for all $r \in \R$ and $\vx \in \R^\Gamma$
  \begin{align*}
   \vf(\vone + r \vx) & = \vf(\vone) + r \vf'(\vone) \vx + r^2 Q(\vx,\vx) && \text{(Taylor expansion)} \\
                      & = \vone + r \vf'(\vone) \vx + r^2 Q(\vx,\vx) && \text{(as $\vf(\vone) = \vone$)} \,.
  \end{align*}
 For $\vu = \vf'(\vone) \vu + \vone$ it follows
  \[
   \vf(\vone + r\vu)  = \vone + r(\vu - \vone) + r^2 Q(\vu,\vu)\,,
  \]
  so we have $\vf(\vone + r\vu) \le \vone + r\vu$ iff $r Q(\vu,\vu) \le \vone$.
 The statement follows.
\qed

\vspace{2mm}
\noindent Next we show how a suitable~$\vw$ can be found.

\begin{lemma} \label{lem:finding-a-w}
 One can compute in polynomial time a vector $\vw \in (1,\infty)^\Gamma$ with \mbox{$\vf(\vw) \ge \vw$.}
\end{lemma}

\begin{proof}
 Using the Taylor expansion of $\vf(\vone + r\vx)$ as in the previous lemma,
  we obtain $\vf(\vone + r\vx) \ge \vone + r\vx$ iff
 \begin{equation} \label{eq:proof-find-a-w}
  r Q(\vx,\vx) \ge (\Id - \vf'(\vone)) \vx\,.
 \end{equation}
 We will choose $\vw := \vone + r\vx$, so we need to find suitable $r$ and $\vx$ such that \eqref{eq:proof-find-a-w} holds.
 Define $\vy \in \{0,1\}^\Gamma$ such that $\vy_X = 1$ if the $X$-component of $Q(\vx,\vx)$ is not constant zero
  (or, equivalently, if there is a rule $X \btran{p} \langle Y,Z\rangle$ for some $Y,Z \in \Gamma$).
 Otherwise, i.e., if $\vf_X(\vx)$ has degree~$1$, set $\vy_X = 0$.
 Define $\vx := \vf'(\vone)^* \vy = (\Id - \vf'(\vone))^{-1} \vy$.
 By the compactness of the task system, all types can reach a type~$X$ with $\vy_X = 1$.
 It follows that $\vf'(\vone)^* \vy$ is positive in all components.
 Hence, $\vx_{\it min} > 0$ where $\vx_{\it min}$ is the smallest component of~$\vx$.

 Observe that $(\Id - \vf'(\vone)) \vx = \vy$, so \eqref{eq:proof-find-a-w} holds at least for the components~$X$ with $\vy_X = 0$.
 Let $c$ denote the smallest nonzero coefficient of~$\vf$.
 Equation~\eqref{eq:proof-find-a-w} holds also for the components~$X$ with $\vy_X = 1$ if
  we set $r > 1 / (c \cdot \vx_{\it min})$.
 The statement follows.
\qed
\end{proof}

\noindent To complete the proof of Theorem~\ref{thm:online} it remains to show
the claimed bounds on~$\pr{S^\sigma \ge k}$.

\begin{qtheorem}{\ref{thm:online}}
 Let $\Delta$ be subcritical.
 \begin{itemize}
  \item
   Let $\vv, \vw \in (1, \infty)^\Gamma$ be vectors with $\vf(\vv) \le \vv$ and $\vf(\vw) \ge \vw$.
   Denote by $\vvmin$ and $\vwmax$ the least component of~$\vv$ and the greatest component of~$\vw$, respectively.
   Then
   \[
     \frac{\vw_{X_0} - 1}{\vwmax^{k+2} - 1} \le \pr{S^\sigma \ge k}
     \le \frac{\vv_{X_0} - 1}{\vvmin^k - 1} \text{ for all online schedulers~$\sigma$.}
   \]
  \item
   Vectors $\vv, \vw \in (1, \infty)^\Gamma$ with $\vf(\vv) \le \vv$ and $\vf(\vw) \ge \vw$ exist and can be computed in polynomial time.
 \end{itemize}
\end{qtheorem}
\begin{proof}
 The second assertion follows from Lemmas \ref{lem:finding-a-v} and~\ref{lem:finding-a-w}.
 It remains to show the first assertion.
 
 Let $h > 1$ and $\vu \in (0,\infty)^\Gamma$ such that $h^{\vu_Y} = \vv_Y$ for all $Y \in \Gamma$.
 Define $\ms{i} := \zs{i} \thickdot \vu$ where ``$\mathord{\thickdot}$'' denotes the scalar product.
 Not that $\ms{1} = \vu_{X_0}$.

 Let us consider $i\geq 1$. Let $y=\css{1},\cdots, \css{i}$ be a sequence
 of elements of $\N^{\Gamma}$ with $\css{i} \ne \vzero$, and let $T_y$ be the set of all
 family trees $t$ satisfying $\zs{j}(t)=\css{j}$ for every $1\leq j\leq i$.
 Note that $\ms{i}(t)\not = 0$.
 Observe that $\ms{i}$ is constant over $T_y$, we denote by $\ms{i}(T_y)$ its value over~$T_y$.

 An easy computation reveals that
 for $Y:=\fsch_{\sigma}(y)$ we have
 \begin{align}\label{eq:hup}
 \Ex{h^{\rs{i} \thickdot \vu} \; \middle\vert \; T_y} & = \Ex{\prod_{Z\in\Gamma}
 h^{\vu_Z \cdot \rs{i}_Z} \; \middle\vert \; T_y} = \Ex{\prod_{Z\in\Gamma}
 \vv_Z^{\rs{i}_Z} \; \middle\vert \; T_y} = \vf_Y(\vv)
  \le \vv_Y = h^{\vu_Y}\,,
  \end{align}
 as $\vf(\vv) \le \vv$.
 Consequently, we have
  \begin{align*}
    \Ex{h^{\ms{i+1}} \mid T_y} 
     & = \Ex{h^{\zs{i+1} \thickdot \vu} \mid T_y}        && \text{(def.~of $\ms{i+1}$)} \\
     & = \Ex{h^{(\zs{i} +\rs{i}- \langle \fsch_{\sigma}(y)\rangle) \thickdot \vu}\mid T_y} &&
       \text{(def.~of $\rs{i}$)} \\
     & = \Ex{h^{\zs{i} \thickdot \vu}\mid T_y}\cdot\Ex{h^{\rs{i}}\mid T_y}\cdot
         \Ex{h^{-\langle \fsch_{\sigma}(y)\rangle \thickdot \vu}\mid T_y} &&
         \!\!\left(\parbox{35mm}{$h^{\zs{i} \thickdot \vu}$, $h^{-\langle \fsch_{\sigma}(y)\rangle \thickdot \vu}$
                const. on $T_y$} \right) \\
     & = h^{\ms{i}(T_y)} \cdot \Ex{h^{\rs{i} \thickdot \vu}\mid T_y}
          \cdot h^{-\vu_Y}&& \text{(def.~of $\ms{i}$)\,.} \\
     & \le h^{\ms{i}(T_y)} && \text{(Equation~(\ref{eq:hup}))}
  \end{align*}
 As this is true for all online schedulers~$\sigma$ and also
 $\Ex{\ms{i+1} \mid \ms{i} = 0} = 0$ we have
 \[
  \Ex{h^{\ms{i+1}} \;\middle\vert\; h^{\ms{1}}, \ldots, h^{\ms{i}}} \le h^{\ms{i}} \,,
 \]
 i.e., the sequence $h^{\ms{1}}, h^{\ms{2}}, \ldots$ is a supermartingale.

Define the stopping time $\tau_k := \inf\{i\ge 1 \mid \ms{i} \in \{0\} \cup [k,\infty)\}$.
Note that $\ms{\tau_k} \le k+2\vumax$, and hence that
$\ms{\tau_k}\in\{0\}\cup [k,k+2\vumax]$.
We wish to apply Doob's Optional-Stopping Theorem~\cite{book:Williams} (sometimes called Optional-Sampling Theorem)
 to infer that $\Ex{h^{\ms{\tau_k}}} \le \Ex{h^{\ms{1}}} = \vv_{X_0}$.
To this end we define the sequence $\hms{1}, \hms{2}, \ldots$ by setting
 $\hms{i} := \ms{i}$ for $i \le \tau_k$ and $\hms{i} := \ms{\tau_k}$ for $i \ge \tau_k$.
The sequence $h^{\hms{1}}, h^{\hms{2}}, \ldots$ is a martingale as $h^{\ms{1}}, h^{\ms{2}}, \ldots$ is a martingale.
To apply the Optional-Stopping Theorem we also need to make sure that $|h^{\hms{i+1}} - h^{\hms{i}}|$
 is bounded by a constant, which is the case as $\hms{i} \in [0,k+2\vumax]$ for all~$i$.
Define the stopping time
  $\tau_k := \inf\{i\ge 1 \mid \ms{i} \in \{0\} \cup [k,\infty)\}$.
 Doob's Optional-Stopping Theorem now yields
 \[
  \Ex{h^{\ms{\tau_k}}} = \Ex{h^{\hms{\tau_k}}} \le \Ex{h^{\hms{1}}} = \Ex{h^{\ms{1}}} = h^{\vu_{X_0}} = \vv_{X_0}\,.
 \]
 Let, as an abbreviation, $p_k := \pr{\ms{\tau_k} \ge k}$.
 Then we have
 \[
  \vv_{X_0} \ge \Ex{h^{\ms{\tau_k}}} \ge h^0 \cdot (1 - p_k) + h^k \cdot p_k = 1 - p_k + h^k \cdot p_k
 \]
 which gives
 \[
  p_k \le \frac{\vv_{X_0} - 1}{h^k - 1} \,.
 \]
 Letting $|\zs{i}|$ denote the sum of the components of~$\zs{i}$, and $\vu_{\mathit min}$ the smallest component of~$\vu$, we have
 \begin{equation}
  \pr{S^\sigma \ge k} = \pr{\sup_i |\zs{i}| \ge k} \le \pr{\sup_i \ms{i} \ge k \vu_{\mathit min}} = p_{k \vu_{\mathit min}} \le \frac{\vv_{X_0} - 1}{\vvmin - 1} \,.
  \label{eq:online-upper-last-line}
 \end{equation}
 So we have shown the upper bound.

 For the lower bound we redefine $h$ and $\vu$ such that $h^{\vu_Y} = \vw_Y$ for all $Y \in \Gamma$ which allows to show in an analogous way that
 \[
  \Ex{h^{\ms{i+1}} \mid h^{\ms{1}}, \ldots, h^{\ms{i}}} \ge h^{\ms{i}} \,,
 \]
 i.e., the sequence $h^{\ms{1}}, h^{\ms{2}}, \ldots$ is now a submartingale.
 The Optional-Stopping Theorem now yields $\Ex{h^{\ms{\tau_k}}} \ge \vw_{X_0}$.
 Further we now have
 \[
  \vw_{X_0} \le \Ex{h^{\ms{\tau_k}}} \le h^0 \cdot (1 - p_k) + h^{k+2\vumax} \cdot p_k = 1 - p_k + h^{k + 2\vumax} \cdot p_k
 \]
 which gives
 \[
  p_k \ge \frac{\vw_{X_0} - 1}{h^{k+2\vumax} - 1}
 \]
 and thus
 \[
  \pr{S^\sigma \ge k} = \pr{\sup_i |\zs{i}| \ge k} \ge \pr{\sup_i \ms{i} \ge k \vumax} = p_{k \vumax} \ge \frac{\vw_{X_0} - 1}{\vwmax^{k+2} - 1} \,.
 \]
\qed
\end{proof}

 \subsection{Proof of Theorem~\ref{thm:light-first}}
\noindent We first prove the following proposition.
\begin{proposition} \label{prop:accumulating-in-P}
The set of $\vv$-accumulating types can be computed in polynomial time.
\end{proposition}
\begin{proof}
We start with some notations.
By $\mathord{\Rightarrow^*}$ we denote the reflexive and transitive closure of~$\mathord{\Rightarrow}$.
We use ``$\mathord{+}$'' for multiset union.
We say that $X$ {\em can generate} a multiset $\alpha$,
denoted by $X \undder{} \alpha$, if some multiset containing $\alpha$ can be derived from
$X$ , i.e., if $X \Rightarrow^* \alpha + \beta$ for some multiset $\beta$.
We write $Y \undder{X} \alpha$ if $Y$ can generate $\alpha$
using only $X$-bounded rules, i.e., rules $Z \hookrightarrow \beta$ such that
$Z \leq X$, and $Y \undder{{\it lf}} \alpha$ to denote that the light-first scheduler can
generate $\alpha$. Finally, we denote by $\alpha^{\geq X}$ ($\alpha^{>X})$
the restriction of
$\alpha$ to types $Y \geq X$ ($Y > X$).

We prove the following characterization: $X$ is $\vv$-accumulating iff there
is $Y$ such that $X_0 \undder{} Y$ and $Y \undder{Y} X +Y$.
This immediately leads to a polynomial algorithm.

($\Rightarrow$): Assume $X$ is $\vv$-accumulating. Then
$X_0 \undder{{\it lf}} n \cdot X$ holds for infinitely many $n \geq 1$.
We claim that there exists a type $W$ such that $W \undder{X} n\cdot X$
for infinitely many $n \geq 1$. For the claim,
take the longest suffixes of the witnesses for
$X_0 \undder{{\it lf}} n\cdot X$ that only use rules $X$-bounded rules,
and let $\alpha_n$ be their corresponding
initial multisets. These suffixes are then witnesses for
$\alpha_n \undder{X} n \cdot X$. By the maximality of the suffixes, either
$\alpha_n = X_0$
holds for infinitely many $n\geq 1$, or
$\alpha_n = \alpha_n^{\geq X}$ does. In the first case, we take $W:= X_0$.
In the second case, let $Z_n \hookrightarrow \beta_n$ be the rule applied to obtain $\alpha_n$. Then
$$X_0 \norarr{{\it lf}}^* (\alpha_n-\beta_n)+ Z_n \norarr{{\it lf}} (\alpha_n - \beta_n)+\beta_n \undder{X} n \cdot X$$
\noindent where $X < Z_n$. Since the step
$(\alpha_n-\beta_n)+ Z_n \norarr{{\it lf}} (\alpha_n - \beta_n)+\beta_n$ is light-first and $X < Z_n$,
we have $(\alpha_n-\beta_n)= (\alpha_n-\beta_n)^{>X}$, and so there are infinitely many $n \geq 1$ such that
$\beta_n \undder{X} n \cdot X$. Since $|\beta_n|\leq 2$ for all $n$,
the type $W$ exists, and the claim is proved.

Consider now a witness of $W \undder{X} n \cdot X$ for some
$n \geq  2^{k}+1$, where $k$ is the number of types. The corresponding tree has depth
at least $k+1$, and so it contains a path in which some type $Y$ appears twice. This easily leads to
$Y \undder{X} X + Y$ for some type $Y$ such that $X_0 \undder{} Y$.

($\Leftarrow$): We start with some simple properties of the
relations $\norarr{X}^*$ and $\norarr{{\it lf}}^*$.
\begin{itemize}
\item[(1)] If $Y \undder{X} \alpha$ and
$\alpha = \alpha^{\geq X}$, then $Y \undder{{\it lf}} \alpha$.\\
Consider a family tree having a (prefix of a) derivation that witnesses $Y \undder{X} \alpha$.
So all ancestors of the nodes corresponding to~$\alpha$ are labeled by symbols that are $\le X$.
It follows that a light-first scheduler may select all ancestors of the $\alpha$-nodes before
 selecting any $\alpha$-node.
Hence $Y \undder{{\it lf}} \alpha$.

\item[(2)] If $X \undder{} Y$ and $Y \undder{{\it lf}} \beta$, then $X \undder{{\it lf}} \beta$.\\
$X \undder{} Y$ implies $X \norarr{\it lf}^* Y + \alpha$ for some~$\alpha$,
 and $Y \undder{{\it lf}} \beta$ implies $Y \norarr{\it lf}^* \beta + \beta_1$ for some~$\beta_1$.
As $X \norarr{\it lf}^* Y + \alpha$, it suffices to find a derivation
 witnessing $Y + \alpha  \norarr{\it lf}^* \emptyset$
 that reaches a multiset of the form $\beta + \gamma$ for some $\gamma$.
Such a derivation is obtained by interleaving the
witnesses for  $Y \norarr{\it lf}^* \beta + \beta_1 \norarr{\it lf}^* \emptyset$ and
$\alpha \norarr{\it lf}^* \emptyset$.
\end{itemize}

Assume now that $X_0 \undder{} Y$ and $Y \undder{X} X + Y$ hold.
Then $Y \undder{X} n \cdot X$ for every $n \geq 1$.
Now (1) yields $Y \undder{{\it lf}} n \cdot X$, and (2) leads to
$X_0 \undder{{\it lf}} n \cdot X$, also for every $n \geq 1$. So
$X$ is $\vv$-accumulating.
\qed
\end{proof}

\noindent Now we complete the proof of Theorem~\ref{thm:light-first}.

\begin{qtheorem}{\ref{thm:light-first}}
 Let $\Delta$ be subcritical and $\vv \in (1, \infty)^\Gamma$ with $\vf(\vv) \le \vv$.
 Let $\sigma$ be a $\vv$-light-first scheduler.
 Let $\vvminmax := \min_{X \btran{} \langle Y, Z \rangle} \max \{ \vv_Y, \vv_Z \}$
 (here the minimum is taken over all transition rules with two types on the right hand side).
 Then $\vvminmax \ge \vvmin$ and for all~$k \ge 1$
 \[
    \pr{S^\sigma \ge k} \le \frac{\vv_{\Init} - 1}{\vvmin \vvminmax^{k-1} - 1}\,.
 \]

 Moreover, let $\vvminacc := \min \{ \vv_X \mid X \in \Gamma,\ X \text{is $\vv$-accumulating} \}$.
 Then $\vvminacc \ge \vvminmax$, $\vvminacc$ can be computed in polynomial time, and there is an integer~$\ell$ such that for all~$k \ge \ell$
 \[
    \pr{S^\sigma \ge k} \le \frac{\vv_{\Init} - 1}{\vvmin^\ell \vvminacc^{k-\ell} - 1} \,.
 \]
\end{qtheorem}
\begin{proof}
 \newcommand{\Li}{\mathit{Li}}
 The inequality $\vvminmax \ge \vvmin$ is trivial.
 For the inequality $\vvminacc \ge \vvminmax$,
  let $\Li := \{Y \in \Gamma \mid \vv_Y < \vvminmax\}$ be the set of types that are strictly lighter than~$\vvminmax$.
 We claim that, in each step~$i$, there is at most one task of $\Li$-type.
 More formally, if $\es{\Li}$ denotes the vector with $\es{\Li}_Y = 1$ for $Y \in \Li$ and
  $\es{\Li}_Y = 0$ for $Y \not\in \Li$, then we have $\zs{i} \thickdot \es{\Li} \le 1$ for all~$i$.
 This can be shown by a straightforward induction on the derivation length:
  at each step the task of $\Li$-type (if present) is selected and replaced by at most two tasks.
 By definition of $\vvminmax$, at most one of the new tasks has $\Li$-type.
 Hence, the types in~$\Li$ are not accumulating.
 It follows $\vvminacc \ge \vvminmax$.

 The rest of the proof is obtained by a small modification of the proof of Theorem~\ref{thm:online}:
 it suffices to show that, in Equation~\eqref{eq:online-upper-last-line},
  we can replace $k \vumin$ by \mbox{$\vumin + (k-1)\vuminmax$} and by \mbox{$\ell \vumin + (k-\ell)\vuminacc$} for some integer~$\ell$.
 (The values $\vuminmax$ and $\vuminacc$ are defined in the obvious way, i.e.,
  using the $h$ from the proof of Theorem~\ref{thm:online} we have $h^{\vuminmax} = \vvminmax$ and $h^{\vuminacc} = \vvminacc$.)
 So we need to show for the light-first scheduler~$\sigma$ that
  $|\zs{i}| \ge k$ implies both $\ms{i} \ge \vumin + (k-1)\vuminmax$ and $\ms{i} \ge \ell \vumin + (k-\ell)\vuminacc$.

 For the first implication, recall that $\ms{i} = \zs{i} \thickdot \vu$.
 We have argued above that $\zs{i} \thickdot \es{\Li} \le 1$.
 This implies $\ms{i} \ge \vumin + (k-1)\vuminmax$.

 For the second implication, let $\ell'$ be an integer such that $\zs{i}_Y \le \ell'$ for all~$i$ and for all non-accumulating types~$Y$.
 Let $\ell := |\Gamma| \cdot \ell'$.
 Then in each step, there are at most $\ell$ tasks of non-accumulating type.
 This implies $\ms{i} \ge \ell \vumin + (k-\ell)\vuminacc$.
\qed
\end{proof}

 \subsection{Proof of Theorem~\ref{thm:depth-first}}

\noindent In the following we let $M^* := I + M + MM + \cdots$ for any square matrix~$M$.
If $M^*$ converges, then, by basic matrix facts, it equals $(\Id - M)^{-1}$.
Also by basic matrix facts (see e.g.~\cite{book:HornJ}), $M^*$ converges iff the spectral radius of~$M$ is less than~one.

Define for all vectors $\vu, \vv$ the vectors $L(\vu)$ and $Q(\vu,\vv)$ such that for all $X \in \Gamma$
 \[
  L(\vu)_X := \sum_{X \btran{p} Y} p \vu_Y \qquad \text{and} \qquad Q(\vu,\vv)_X := \sum_{X \btran{p} YZ} p \vu_Y \vu_Z\,.
 \]
Note that the sums extend over the rules after applying~$\lambda$.
Also note that $L$ is a linear vector function and we view it as a matrix whose rows and columns are indexed with~$\Gamma$.
Furthermore, we write $Q(\cdot,\vv)$ and $Q(\vu,\cdot)$ for
the matrices with $Q(\cdot,\vv) \vu = Q(\vu,\vv) = Q(\vu, \cdot) \vv$.

\noindent Here is a restatement of Theorem~\ref{thm:depth-first}:
\begin{qtheorem}{\ref{thm:depth-first}}
 \stmtthmdepthfirst
\end{qtheorem}
\vspace{2mm}

We first prove the first part of Theorem~\ref{thm:depth-first}.
In fact, the following proposition allows to compute $\pr{S^\sigma_X \ge k}$ for all $X \in \Gamma$ at the same time.
We define, for all $k \ge 1$, the vector $\vs[k] \in [0,1]^\Gamma$ such that $\vs[k]_X = \pr{S^\sigma_X \ge k}$ for all~$X$.

\begin{proposition} \label{prop:depth-first-distribution}
  Let $A[k] := L + Q(\vone - \vs[k], \cdot)$.
  Then $(I-A[k])^{-1}$ exists and for all $k \ge 1$
  \[
   \vs[k+1] = A[k] \vs[k+1] + Q(\cdot,\vone) \vs[k] = (I-A[k])^{-1} Q(\cdot,\vone) \vs[k]\,.
  \]
\end{proposition}

\begin{proof}
 The following equation follows from the definition of a depth-first scheduler~$\sigma$.
 \begin{align*}
  \pr{S^\sigma_X \ge k+1} & = \sum_{X \btran{p} Y}  p \pr{S^\sigma_Y \ge k+1} \\
                          & \quad + \sum_{X \btran{p} YZ} p \left( \pr{S^\sigma_Y \ge k} + \pr{S^\sigma_Y < k} \cdot \pr{S^\sigma_Z \ge k+1} \right)
 \end{align*}
 Using the definitions this immediately implies the equality
  \[
   \vs[k+1] = A[k] \vs[k+1] + Q(\cdot,\vone) \vs[k]\,.
  \]
 For the second equality of the proposition, note that $\vf'(\vone) = L + Q(\vone,\cdot) + Q(\cdot,\vone)$.
 As the task system is subcritical, the spectral radius of $\vf'(\vone)$ is, by Proposition~\ref{prop:critical-spectral}, less than~one.
 So the spectral radius of $A[k] \le L + Q(\vone,\cdot) \le \vf'(\vone)$ is less than~one as well.
 Hence, by standard matrix facts~\cite{book:HornJ} the sum $A[k]^*$ converges and equals $(I - A[k])^{-1}$.
 The second equality follows.
\qed
\end{proof}
Notice that Proposition~\ref{prop:depth-first-distribution} in fact implies the first statement of Theorem~\ref{thm:depth-first},
 because $\pr{S^\sigma = k} = \vs[k]_{X_0} - \vs[k-1]_{X_0}$ and a matrix can be inverted in time $\bigo(|\Gamma|^3)$ in the unit-cost model.

For the rest of the proof of Theorem~\ref{thm:depth-first} we need the following two auxiliary lemmata.
\begin{lemma} \label{lem:eps-out-of-star}
 Let $A$ be a nonnegative square matrix with spectral radius less than~one.
 Let $(\epsilon_{n})_{n\in\Nat}$ be a sequence with $\epsilon_{n} \ge \epsilon_{n+1} \ge 0$ converging to~$0$.
 Then there exists an $n_1$ and a nonnegative matrix $K$ such that for all $n \ge n_1$
  \[
   \big( (1 - \epsilon_{n}) A \big)^* \ge (I - \epsilon_{n} K) A^* \;.
  \]
\end{lemma}
\begin{proof}
 We can assume $\epsilon_{n} \le 1$.
 Let $M = (I - A)^{-1}A$.
 Then by a simple computation
  \[
   \big( (1 - \epsilon_{n}) A \big)^* = \big( I + \epsilon_{n} M \big)^{-1} A^*\;.
  \]
 Choose $n_1$ large enough so that $\rho(\epsilon_{n} M) < 1$.
 Then $(\epsilon_{n}M)^*$ exists and so
  \begin{eqnarray*}
   \big( I + \epsilon_{n} M \big)^{-1} & = &   I - (\epsilon_{n}M) + (\epsilon_{n}M)^2 - (\epsilon_{n}M)^3 + - \cdots \\
                                       & \ge & I - (\epsilon_{n}M) (\epsilon_{n}M)^* \\
                                       & \ge & I - \epsilon_{n} M (\epsilon_{n_1}M)^*
  \end{eqnarray*}
 Choose $K = M (\epsilon_{n_1} M)^*$ and the claim follows.
\qed
\end{proof}

\begin{lemma} \label{lem:spectral-radius-of-B}
 Let $B := (\Id - L - Q(\vone,\cdot))^{-1} Q(\cdot,\vone)$.
 Then the spectral radius of~$B$ is less than~1.
\end{lemma}
\begin{proof}
 Observe that $\vf'(\vone) = L + Q(\vone,\cdot) + Q(\cdot,\vone)$.
 As $(\Delta,X)$ is subcritical, Proposition~\ref{prop:critical-spectral} implies
  that the spectral radius of~$\vf'(\vone)$ is less than~one.
 Then it follows that the spectral radius of~$B$ is less than one as well,
  using the theory of M-matrices and regular splittings,
  see \cite{book:BermanP}, Theorem~6.2.3 part P$_{48}$.
\qed
\end{proof}

\noindent To complete the proof of Theorem~\ref{thm:depth-first} it suffices to show the following proposition.

\begin{proposition}\label{prop:depth-first-asymptotic}
 Let $\Delta$ be subcritical and $\sigma$ be any depth-first scheduler.
 Let $B := \left( L + Q(\vone, \cdot) \right)^* Q(\cdot,\vone)$ and $\rho$ the spectral radius of~$B$.
 Then $0 < \rho < 1$ and $\pr{S^\sigma \ge k} \in \Theta(\rho^k)$,
  i.e, there are $c,C > 0$ such that $c \rho^k \le \pr{S^\sigma \ge k} \le C \rho^k$ for all~$k$.
\end{proposition}
\begin{proof}
 We have $\rho < 1$ by Lemma~\ref{lem:spectral-radius-of-B}.
 To show $\rho > 0$, it suffices (by Perron-Frobenius theory~\cite{book:BermanP}) to show that
  all row sums of~$B$ are (strictly) positive.
 For this, let $Y \in \Gamma$ be the index of an arbitrary row.
 Then, by compactness of the task system, there are types $X_0, \ldots, X_i$ ($0 \le i \le n-1$) such that
  $Y = X_i$ and $X_i \btran{p_i} X_{i-1}, \ldots, X_1 \btran{p_1} X_0$ and $X_0 \btran{p_0} Z W$
  for some $Z,W \in \Gamma$.
 It is straightforward to show by induction on~$i$ that the $(Y,Z)$-entry of $L^i Q(\cdot,\vone)$ is positive.
 It follows that the $(Y,Z)$-entry of~$B$ is positive, so $\rho > 1$.

 For the upper bound, observe that with Proposition~\ref{prop:depth-first-distribution} we have
  \begin{equation} \label{eq:proof-thm-depth-first-asy1}
   \vs[k+1] = \left( L + Q(\vone - \vs[k], \cdot) \right)^* Q(\cdot,\vone) \vs[k] \le B \vs[k]\,.
  \end{equation}
 By a simple induction it follows $\vs[k+i] \le B^i \vs[k]$.
 As the absolute values of the eigenvalues of~$B$ are bounded by~$\rho$ we get $\norm{\vs[k+i]} \le C_1 \rho^i$ for some $C_1>0$,
  which implies the claimed upper bound.

 For the lower bound, observe that there is a real number $0 < r \le 1$ such that for all types $Y \in \Gamma$,
  the probability that $X$ reaches~$Y$ is at least~$r$.
 So it suffices to find any $Y \in \Gamma$ such that there is a $c_1 > 0$ with $\pr{S^\sigma_Y \ge k} \ge c_1 \rho^k$ for all~$k$.

 \newcommand{\Gunb}{\Gamma_{\uparrow}}
 Recall that $\rho$ is the spectral radius of~$B$.
 It is a corollary (Corollary 2.1.6 of~\cite{book:BermanP}) of Perron-Frobenius theory
  that $B$ has a principal submatrix~$B'$ which is irreducible and also has spectral radius~$\rho$.
 We write $\Gunb$ for the subset of~$\Gamma$ such that $B'$ is obtained from~$B$ by deleting all rows and columns
  that are not indexed by~$\Gunb$.
 Also by Perron-Frobenius theory, $B'$ has an eigenvector $\vu' \in (0,\infty)^{\Gunb}$
  with $B'\vu' = \rho \vu'$ so that $\vu'$ is positive in all components.
 Define $\vu \in [0,\infty)^\Gamma$ as the vector with $\vu_Y = \vu'_Y > 0$ for $Y \in \Gunb$ and $\vu_Y = 0$ for $Y \not\in \Gunb$.
 Hence we have $B \vu \ge \rho \vu$.
 By the already proven upper bound there is a $t > 0$ such that $\vs[k] \le t \rho^k$ for all~$k$.
 We abbreviate $\epsilon_k := t \rho^k$ so that $\vs[k] \le \epsilon_k \vone$.

 Now we show that there is a natural number~$k$ and a real number $d > 0$ with $\epsilon_k d < 1$ such that for all~$i \ge 0$
  \begin{equation} \label{eq:proof-thm-depth-first-ind}
   \vs[k+i] \ge \rho^i \left( \prod_{j=1}^i (1 - \epsilon_{k+j-1}d) \right) \vu\,.
  \end{equation}
 As $\vu_Y = 0$ for $Y \not\in \Gunb$ it suffices to show $\vs[k+i] \ge_\uparrow \rho^i \left( \prod_{j=1}^i (1 - \epsilon_{k+j-1}d) \right) \vu$
  where by the notation $\vv \ge_\uparrow \vw$ we mean $\vv_Y \ge \vw_Y$ for all $Y \in \Gunb$.
 We proceed by induction on~$i$ and determine the constants on the fly.
 For the induction base ($i=0$) observe that, as $\vs[k]$ is positive by compactness of the task system, we can enforce $\vs[k] \ge \vu$
  by scaling down $\vu$ by multiplying it with a small constant.
 This does not affect the stated properties of~$\vu$.
 For the step, let $i \ge 0$.
 We have
 \begin{align*}
  \vs[k+i+1]
    & =   \left( L + Q(\vone - \vs[k+i], \cdot) \right)^* Q(\cdot,\vone) \vs[k+i]
                        && \text{(by~\eqref{eq:proof-thm-depth-first-asy1})} \\
    & \ge \left( (1-\epsilon_{k+i}) (L + Q(\vone,\cdot)) \right)^* Q(\cdot,\vone) \vs[k+i]
                        && \text{(as $\vs[k+i] \le \epsilon_{k+i} \vone$)} \\
    & \ge \left( (1-\epsilon_{k+i}) (L + Q(\vone,\cdot)) \right)^* Q(\cdot,\vone) \rho^i \left( \prod_{j=1}^i (1 - \epsilon_{k+j-1}d) \right) \vu
                        && \text{(ind.~hypothesis)} \\
    & \ge (\Id - \epsilon_{k+i} K) B \rho^i \left( \prod_{j=1}^i (1 - \epsilon_{k+j-1}d) \right) \vu
                        && \left(\parbox{3cm}{for a large~$k$ and some matrix~$K$ by Lemma~\ref{lem:eps-out-of-star}} \right) \\
    & \ge \rho^i \left( \prod_{j=1}^i (1 - \epsilon_{k+j-1}d) \right) (\rho \vu - \epsilon_{k+i} K B \vu)
                        && \text{(as $B \vu \ge \rho \vu$)} \\
    & \ge_\uparrow \rho^i \left( \prod_{j=1}^i (1 - \epsilon_{k+j-1}d) \right) (\rho \vu - \epsilon_{k+i} \rho d \vu)
                        && \left(\parbox{3cm}{for a large $d$ with $K B \vu \le_\uparrow \rho d \vu$} \right) \\
    & = \rho^{i+1} \left( \prod_{j=1}^{i+1} (1 - \epsilon_{k+j-1}d) \right) \vu
 \end{align*}
 This proves~\eqref{eq:proof-thm-depth-first-ind}.
 So, denoting by $\vu_{\mathit min} > 0$ the smallest nonzero component of~$\vu$, we have
  \[ \vs[k+i]_Y \ge \rho^i \left( \prod_{j=1}^{i+1} (1 - \epsilon_{k+j-1}d) \right) \vu_{\mathit min} \qquad \text{for all $Y \in \Gunb$ and all $i \ge 0$.}
  \]
 Thus the proof is completed if $\prod_{j=k}^\infty (1 - \epsilon_{j} d) > 0$.
 To see that this inequality holds, observe that $1 - \epsilon_j d = 1 - t \rho^j d \ge 1 - \frac{1}{j^2}$ is true for almost all~$j$
  and that $\prod_{j=2}^\infty (1 - \frac{1}{j^2}) = \frac{1}{2} > 0$.
 This completes the proof.
\qed
\end{proof}

 \section{Proofs of Section~\ref{sec:expectations}} \label{sec:app-expectations}

\subsection{Proof of Theorem~\ref{thm:whitebox-exp}}
\begin{qtheorem}{\ref{thm:whitebox-exp}}
The expectation $\Ex{\xo}$ is finite (no matter whether $\Delta$ is
critical or subcritical). Moreover, $\bigo(b)$ terms compute $b$ bits of
$\Ex{\xo}$. If the task system $\Delta$ is subcritical, then $\log_2 b + \bigo(1)$
terms compute $b$ bits of $\Ex{\xo}$.
Finally, computing $k$ terms takes time $\bigo(k \cdot |\Gamma|^3)$ in the unit cost model.
\end{qtheorem}
\begin{proof}
Note that the second statement implies the first one.
Let $\esc{i} := 1 - \ns{i}_\Init$.
Then we have $\Ex{\xo} - \sum_{i=0}^{k-1} (1 - \ns{i}_\Init) = \sum_{i=k}^\infty \esc{i}$.
It follows from \cite{EKL08:stacs} that there is a $c_1 \in (0,\infty)$
 such that for all $i \in \N$ we have $\esc{i} \le c_1 \cdot 2^{-i/(n2^n)}$ where $n = |\Gamma|$.
Using this inequality we get
 \[
  \sum_{i=k}^\infty \esc{i} \le c_1 \sum_{i=k}^\infty 2^{-i/(n2^n)} \le c_2 \cdot 2^{-k/(n2^n)}
 \]
 with $c_2 = c_1 / (1 - 2^{-1/(n2^n)})$.
Choosing $k = \lceil (b + \log_2 c_2) n 2^n \rceil$ we obtain $\sum_{i=k}^\infty \esc{i} \le 2^{-b}$ which proves the second statement.

For the third statement (about subcritical systems) recall from Corollary~\ref{cor:optimal-expectation} that
 there are $c > 0$ and $0 < d < 1$ such that $\esc{i} \le c \cdot d^{2^i}$ for all $i \in \Nat$.
So
 \[ \sum_{i=k}^\infty \esc{i} \le \sum_{i=k}^\infty c \cdot d^{2^i} \le c \cdot \sum_{i=0}^\infty d^{2^k + i} = \frac{c}{1-d} \cdot d^{2^k} \,.
 \]
By choosing a natural number $k$ with $k \ge -\log_2(-\log_2 d) + \log_2 b + 1$ we obtain for all $b \ge \log \frac{c}{1-d}$ that
 $\frac{c}{1-d} \cdot d^{2^k} \le 2^{-b}$ which proves the third statement.

The final statement follows from Corollary~\ref{cor:optimal-distribution-cost}.
\qed
\end{proof}

\subsection{Proof of Theorem~\ref{thm:online-equivalence}}
\begin{qtheorem}{\ref{thm:online-equivalence}}
If $\Delta$ is subcritical, then $\Ex{S^{\sigma}}$ is finite for every online scheduler~$\sigma$.
If $\Delta$ is critical, then $\Ex{S^{\sigma}}$ is infinite for every online
scheduler~$\sigma$.
\end{qtheorem}
\begin{proof}
Let $\Delta$ be subcritical.
By Theorem~\ref{thm:online} we have for every online scheduler~$\sigma$
 \[
  \Ex{S^{\sigma}}
    = \sum_{k=1}^\infty \pr{S^{\sigma} \ge k}
    \le \sum_{k=1}^\infty \frac{\vv_{X_0} - 1}{\vvmin^k - 1} < \infty\,,
 \]
 because it is a geometric series.
 
Let now $\Delta$ be critical.
The proof follows the lines of the proof of Theorem~\ref{thm:online}.
By Proposition~\ref{prop:critical-spectral} we have $\rho(\vf'(\vone)) = 1$
 for the spectral radius of~$\vf'(\vone)$.

Let us fix an online scheduler $\sigma$.
First we prove $\Ex{S^{\sigma}} = \infty$ for the case in which
$X_0$ is reachable from every type $X \in \Gamma$.
Later we will show how to drop this assumption.
If $X_0$ is reachable from every~$X$, it follows that $\vf'(\vone)$
 is an irreducible matrix.
Then Perron-Frobenius theory \cite{book:BermanP} guarantees the existence
of an eigenvector $\vu \in \R^\Gamma$ of $\vf'(\vone)$ which
 is positive in all components, i.e., $\vf'(\vone) \vu = \vu$ and
 $\vu_X > 0$ for all $X \in \Gamma$.
W.l.o.g.\ we can choose $\vu$ such that its largest component is~$1$.
Let again $\ms{i} := \zs{i} \thickdot \vu$.
Note that $\ms{1} = \vu_{X_0} > 0$ and $\ms{i} \le |\zs{i}|$ where $|\zs{i}|$ denotes the sum of the components of~$\zs{i}$.
Also note that $\ms{i}$ returns a weighted sum of the components of~$\zs{i}$.
Loosely speaking, we will show that its expectation remains constant.

Let us consider $i\geq 1$. Let $y=\css{1},\cdots, \css{i}$ be a sequence
of elements of $\N^{\Gamma}$ with $\css{i} \ne \vzero$, and let $T_y$ be the set of all
family trees $t$ satisfying $\zs{j}(t)=\css{j}$ for every $1\leq j\leq i$.
Note that $\ms{i}(t)\not = 0$.
Observe that $\ms{i}$ is constant over $T_y$, we denote by $\ms{i}(T_y)$ its value over~$T_y$.


 An easy computation reveals that
 for every $X \in \Gamma$ we have
\[
    \Ex{\rs{i}_X \mid T_y}
  = \sum_{\fsch_{\sigma}(y)\btran{p} \alpha} p\cdot \#_X(\alpha)
  = \vf_{\fsch_{\sigma}(y), X}'(\vone)
\]
 which gives
 \begin{equation} \label{eq:proof-online-equiv-mean}
 \Ex{\rs{i}\mid T_y}=\vf_{\fsch_{\sigma}(y)}'(\vone)
 \end{equation}
 (where $\vf_{\fsch_{\sigma}(y)}'(\vone)$ denotes the row vector indexed by~${\fsch_{\sigma}(y)}$).
 Consequently, we have:
 \begin{align*}
  \Ex{\ms{i+1} \mid T_y}
    & = \Ex{\zs{i+1} \mid T_y} \thickdot \vu   &&
    \text{(def.~of $\ms{i+1}$)} \\
    & = \left( \Ex{\zs{i}\mid T_y} +
        \Ex{\rs{i}\mid T_y} - \Ex{\langle \Xs{i}\rangle\mid T_y}\right)
    \thickdot \vu  && \text{(def.~of $\rs{i}$)} \\
    & = \left( \Ex{\zs{i}\mid T_y} +
        \vf_{\fsch_{\sigma}(y)}'(\vone) - \langle \fsch_{\sigma}(y)\rangle
         \right)
    \thickdot \vu && \text{(by~\eqref{eq:proof-online-equiv-mean})} \\
    & = \ms{i}(T_y) + \vf_{\fsch_{\sigma}(y)}'(\vone)\vu -
    \langle \fsch_{\sigma}(y)\rangle \thickdot \vu && \text{(def.~of $\ms{i}(T_y)$)} \\
    & = \ms{i}(T_y)
    && \text{(as $\vf'(\vone) \vu = \vu$)}
 \end{align*}
Also clearly $\Ex{\ms{i+1} \mid \ms{i} = 0} = 0$, and hence we have
 \[
  \Ex{\ms{i+1} \mid \ms{1}, \ldots, \ms{i}} = \ms{i} \,,
 \]
 i.e., the sequence $\ms{1}, \ms{2}, \ldots$ is a martingale.

Define the stopping time $\tau_k := \inf\{i\ge 1 \mid \ms{i} \in \{0\} \cup [k,\infty)\}$.
Note that $\ms{\tau_k} \le k+2$ as $\vu \le \vone$, and hence that
$\ms{\tau_k}\in\{0\}\cup [k,k+2]$.
We wish to apply Doob's Optional-Stopping Theorem~\cite{book:Williams} (sometimes called Optional-Sampling Theorem)
 to infer that $\Ex{\ms{\tau_k}} = \Ex{\ms{1}} = \vu_{X_0}$.
To this end we define the sequence $\hms{1}, \hms{2}, \ldots$ by setting
 $\hms{i} := \ms{i}$ for $i \le \tau_k$ and $\hms{i} := \ms{\tau_k}$ for $i \ge \tau_k$.
The sequence $\hms{1}, \hms{2}, \ldots$ is a martingale as $\ms{1}, \ms{2},
\ldots$ is a martingale.
To apply the Optional-Stopping Theorem we also need to make sure that $|\hms{i+1} - \hms{i}|$
 is bounded by a constant, which is the case as $\hms{i} \in [0,k+2]$
  for all~$i$.
%
Doob's Optional-Stopping Theorem now yields
 \[
  \Ex{\ms{\tau_k}} = \Ex{\hms{\tau_k}} = \Ex{\hms{1}} = \vu_{X_0}\,.
 \]
Recall that this is~$>0$.
Since $\ms{\tau_k}\in\{0\}\cup [k,k+2]$,
 \[
  \vu_{X_0} = \Ex{\ms{\tau_k}} \le 0 \cdot
  \pr{\ms{\tau_k}=0} + (k+2) \cdot \pr{\ms{\tau_k} \ge k}=
  (k+2) \cdot \pr{\ms{\tau_k} \ge k}
 \]
which gives
 \[
  \pr{\ms{\tau_k} \ge k} \ge \frac{\vu_{X_0}}{k+2} \,.
 \]
So we have
 \begin{align*}
  \pr{S^{\sigma} \ge k} = \pr{\sup_i |\zs{i}| \ge k} \ge \pr{\sup_i \ms{i} \ge k} = \pr{\ms{\tau_k} \ge k} \ge \frac{\vu_{X_0}}{k+2} \,.
 \end{align*}
Hence,
 \[
  \Ex{S^{\sigma}} = \sum_{k=1}^\infty \pr{S^{\sigma} \ge k}
  \ge \sum_{k=1}^\infty \frac{\vu_{X_0}}{k+2} = \infty
 \]
 which completes the proof for the case where $X_0$ is reachable from all types.

Now we show that $\Ex{S^{\sigma}} = \infty$ also holds when
$X_0$ is not reachable from all types.
Recall that $\rho(\vf'(\vone)) = 1$.
It is a corollary (Corollary 2.1.6 of~\cite{book:BermanP}) of Perron-Frobenius theory
 that $\vf'(\vone)$ has a principal submatrix~$B$ which is irreducible and has spectral radius $\rho(B) = 1$.
Let $\Gamma' \subseteq \Gamma$ denote the set of types such that $B$ is obtained from $\vf'(\vone)$
 by deleting all rows and columns not indexed by $\Gamma'$.
Consider the task system $\Delta'$ which is the original task system
restricted to~$\Gamma'$.
More concretely, $\Delta'$ has types $\Gamma'$ and transition rules as follows:
A rule $X \btran{p} \alpha'$ is in $\Delta'$ iff $X \in \Gamma'$ and
 there is an $\alpha \in M_\Gamma^{\le 2}$ such that $X \btran{p} \alpha$ is
 in the original task system 
 and $\alpha'$ is obtained from $\alpha$ by deleting the types that are not in~$\Gamma'$.
Let $\vg: \R^{\Gamma'} \to \R^{\Gamma'}$ denote the pgf for~$\Delta'$.
From the construction of~$\Delta'$ it is straightforward to see that
$B = \vg'(\vone)$.
Pick an arbitrary $X \in \Gamma'$ as the initial type of $\Delta'$.
As $B = \vg'(\vone)$ is irreducible, $X$ is reachable from all types
in~$\Gamma'$.
Hence, the first part of the proof applies and we obtain that,
 in $\Delta'$, we have $\Ex{S^{\sigma}_X} = \infty$ for all
 online schedulers~$\sigma$.
As $\Delta'$ was obtained by erasing types and rules from the original
task system,
 it is easy to see that, also in the original task system,
 we have $\Ex{S^{\sigma}_X} = \infty$ for all online schedulers~$\sigma$.
As $X$ is reachable from~$X_0$, it follows $\Ex{S^{\sigma}} = \infty$
for all online schedulers~$\sigma$.
\qed
\end{proof}

}{}

\end{document}